\documentclass[submission, Phys]{SciPost}
\hypersetup{
    colorlinks,
    linkcolor={red!50!black},
    citecolor={blue!50!black},
    urlcolor={blue!80!black}
}

\usepackage[bitstream-charter]{mathdesign}
\DeclareSymbolFont{usualmathcal}{OMS}{cmsy}{m}{n}
\DeclareSymbolFontAlphabet{\mathcal}{usualmathcal}

\usepackage{ulem}
\usepackage{bbm}
\usepackage{multirow}
\usepackage{color}
\usepackage{hyperref}
\usepackage{braket}
\usepackage[justification=justified]{caption}
\usepackage{subcaption}
\graphicspath{ {./} {./FIGS/} }
\begin{document}

% TODO: write your article's title here.
% The article title is centered, Large boldface, and should fit in two lines
\begin{center}{\Large \textbf{Quantum order-by-disorder induced phase transition in Rydberg
ladders with staggered detuning
      \\
}}\end{center}
%old title: Disorder-free localization via Hilbert space fragmentation in two dimensions
% TODO: write the author list here. Use first name (+ other initials) + surname format.
% Separate subsequent authors by a comma, omit comma and use "and" for the last author.
% Mark the corresponding author with a superscript star.
\begin{center}
Madhumita Sarkar, Mainak Pal, Arnab Sen, and K. Sengupta
\end{center}

% TODO: write all affiliations here.
% Format: institute, city, country
\begin{center}
School of Physical Sciences, Indian Association for the Cultivation
of Science, Kolkata 700032, India
%\\
%{\bf 2} Department of Physics and Astronomy, University College
%London, Gower Street, London WC1E 6BT, United Kingdom
\\
% TODO: provide email address of corresponding author
${}^\star$ {\small \sf ksengupta1@gmail.com}
\end{center}

\begin{center}
\today
\end{center}

% For convenience during refereeing (optional),
% you can turn on line numbers by uncommenting the next line:
%\linenumbers
% You should run LaTeX twice in order for the line numbers to appear.

\section*{Abstract}

{\bf $^{87}{\rm Rb}$ atoms are known to have long-lived Rydberg
excited states with controllable excitation amplitude (detuning) and
strong repulsive van der Waals interaction $V_{{\bf r} {\bf r'}}$
between excited atoms at sites ${\bf r}$ and ${\bf r'}$. Here we
study such atoms in a two-leg ladder geometry in the presence of
both staggered and uniform detuning with amplitudes $\Delta$ and
$\lambda$ respectively. We show that when $V_{{\bf r r'}} \gg(\ll)
\Delta, \lambda$ for $|{\bf r}-{\bf r'}|=1(>1)$, these ladders host
a plateau for a wide range of $\lambda/\Delta$ where the ground
states are selected by a quantum order-by-disorder mechanism from a
macroscopically degenerate manifold of Fock states with fixed
Rydberg excitation density $1/4$. Our study further unravels the
presence of an emergent Ising transition stabilized via the
order-by-disorder mechanism inside the plateau. We identify the
competing terms responsible for the transition and estimate a
critical detuning $\lambda_c/\Delta=1/3$ which agrees well with
exact-diagonalization based numerical studies. We also study the
fate of this transition for a realistic interaction potential
$V_{{\bf r} {\bf r'}} = V_0 /|{\bf r}-{\bf r'}|^6$, demonstrate that
it survives for a wide range of $V_0$, and provide analytic estimate
of $\lambda_c$ as a function of $V_0$. This allows for the
possibility of a direct verification of this transition in standard
experiments which we discuss.

}

% Guideline: if your paper is longer that 6 pages, include a TOC
% To remove the TOC, simply cut the following block
\vspace{10pt} \noindent\rule{\textwidth}{1pt}
\tableofcontents\thispagestyle{fancy}
\noindent\rule{\textwidth}{1pt} \vspace{10pt}

\section{Introduction}
\label{int}

It is usually expected that fluctuations, thermal or quantum, in a
generic many-body system shall lead to suppression of order.
However, a somewhat less intuitive counterexample, namely,
stabilization of order in a system with competing interactions due
to quantum or thermal fluctuations is now well-accepted
\cite{villain1,shender1,shenderrev}. This phenomenon occurs due to
the presence of macroscopically degenerate manifold of classical
ground states in such systems; the presence of fluctuations then may
lift this degeneracy leading to a ground state with definite order.
This mechanism is dubbed as order-by-disorder \cite{villain1}.
Examples of this phenomenon is seen in a variety of quantum
many-body systems involving spins \cite{spinref1,spinref2,spinref3},
bosons \cite{bosonref1,bosonref2,bosonref3} and fermions
\cite{fermiref}.

In recent years, ultracold atoms in optical lattice have proved to
be efficient emulators of several well-known model Hamiltonians in
condensed matter systems \cite{blochrev}. A primary example of this
is the emulation of the Bose-Hubbard model
\cite{fisher1,tvr1,jaksch1,nicolas1} using ultracold bosonic atoms
\cite{blochexp}. A study of such bosons in their Mott phases and in
the presence of a tilted lattice generating an artificial electric
field have led to the realization of a translational symmetry broken
ground state and an associated quantum phase transition which
belongs to the Ising universality class \cite{subir1a,subir1b,
kolov1,kolov2, powel1,rey1,mukh1,bakr1}. More recently, another such
ultracold atom system, namely, a chain of ${^{87}}{\rm Rb}$ atoms
supporting long-lived Rydberg excited states has also been studied
in detail. The amplitude of realizing a Rydberg excitation at any
site of such a chain can be controlled by changing the detuning,
{\it i.e}, the difference between the energy (frequency) of the
external laser and the energy gap between the ground and excited
states of a Rydberg atom \cite{rydexp1,rydexp2,rydexp3,rydexp4}.
These atoms, in their Rydberg excited states, experience strong
repulsive van der Waals(vdW) interaction leading to a finite Rydberg
blockade radius \cite{rydbl1,rydbl2,rydbl3,rydbl4,rydbl5}. Such a
system leads to the realization of symmetry broken phases separated
by both Ising and non-Ising quantum phase transitions
\cite{rydexp1,rydexp2,roopayan1,subir2,subir3,other1}. The
out-of-equilibrium dynamics of such systems leading to realization
of the central role of quantum scars in such dynamics have also been
studied both theoretically \cite{ethv01,ethv02,ethv03,ethv04,
ethv05,ethv06} and experimentally \cite{rydexp2}. These studies have
also been extended to two-dimensional (2D) arrays of Rydberg atoms
leading to the possibility of realization of Kitaev spin liquids in
these systems \cite{ryd2d1,ryd2d2,ryd2d3,ryd2d4,ryd2d5,ryd2d6}.

In all of the above-mentioned studies
\cite{rydexp1,rydexp2,rydexp3,rydexp4,roopayan1,subir2,subir3,other1,ryd2d1,ryd2d2,ryd2d3,ryd2d4,ryd2d5,ryd2d6,
ethv01,ethv02,ethv03,ethv04, ethv05,ethv06}, detuning of the Rydberg
atoms have been assumed to be uniform. However, recently, chains of
such Rydberg atoms in the presence of an additional staggered
detuning have also been studied extensively
\cite{simu1,simu2,simu3,simu4,simu5,deb1,deb2,conf1,conf2,bm1}. The
main motivation behind such studies stemmed from the possibility of
realization of quantum link models using ultracold atom systems
\cite{simu1,simu2,simu3,simu4,deb1,deb2}; these models are
well-known to exhibit quantum confinement\cite{qedconf1,qedconf2}
which is usually difficult to realize in a generic condensed matter
setup \cite{conf1,conf2,conf3}. More recently, the
out-of-equilibrium dynamics of such models have also been studied;
it was found that they host several interesting phenomena such as
ultra-slow dynamics following a quench \cite{conf1}, dynamical
freezing in the presence of a periodic drive, and Floquet scars
\cite{bm1}. However, to the best of our knowledge, the phases of
coupled ladders of Rydberg atoms with staggered detuning have not
been studied in detail so far.

\begin{figure}[ht]
\centering
\includegraphics[width=\linewidth]{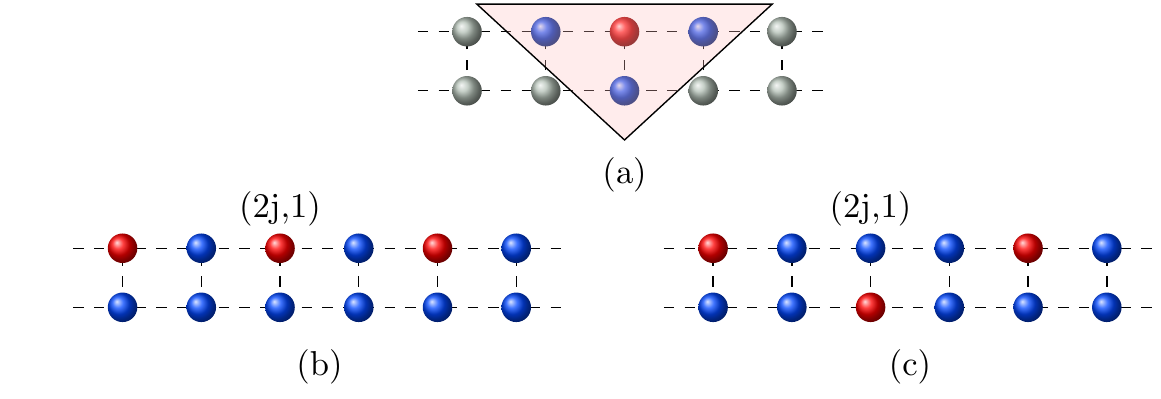}
\caption{(a) Schematic representation of a two-leg Rydberg ladder.
The circles represent sites on the Rydberg chain. The presence of a
Rydberg excitation on a site is denoted by the red circle; this
precludes excitations of nearby sites (denoted by blue circles). The
corresponding Rydberg blockade radius is schematically shown by the
triangle. The other sites (grey circles) may have Rydberg
excitations since they would experience weaker repulsive vdW
interaction falling off as $1/|\vec r|^6$ with distance $|\vec r|$
from the site with a Rydberg excited atom (red circle). (b)
Schematic representation of one of the states with Rydberg
excitation density $n=1/4$ with excitations occurring at even sites
of one of the ladders. This state is one of the two possible
classical Fock states which represents the ground state for
$\lambda>\lambda_c$. (c) A similar diagram for a state in the
low-energy manifold with $n=1/4$ whose superposition forms the
ground state at $\lambda<\lambda_c$ but not for $\lambda>\lambda_c$.
See text for details.} \label{fig0}
\end{figure}

In this work, we study the phases of such coupled chains of Rydberg
atoms in the presence of both uniform and staggered detunings with
amplitudes $\lambda$ and $\Delta$ respectively. In addition, such a
system allows for a coupling between the Rydberg excited and the
ground states of the atoms with an amplitude $w$
\cite{rydexp1,rydexp2}. Such coupled chains leads to a two-leg
ladder geometry as shown in Fig.\ \ref{fig0}(a). The corresponding
blockade radius for the Rydberg excitations, which prevents
existence of two Rydberg excitation on nearest-neighbor sites in
such ladders, is also shown in Fig.\ \ref{fig0}(a). In our work, we
first concentrate in the regime where the vdW interaction between
the Rydberg atoms satisfy $V_{{\bf r} {\bf r'}} \ll \lambda ,\Delta,
w$ outside the blockade radius. In this regime, we find that for a
wide range of $\lambda/\Delta$ such a system shows a plateau where
the Rydberg excitation density, $n$, is fixed to $n = 1/4$. The
plateau destabilizes at $\lambda/\Delta \simeq \pm 1$ leading to a
change in $n$; we analyze this behavior using a variational
wavefunction method which agrees well with exact diagonalization
(ED) results.

The ground state of the system within the plateau involves
macroscopic number of degenerate classical Fock states having a
Rydberg excitation in one of the two sites in every even rung of the
ladder (Fig.\ \ref{fig0}(b) and \ref{fig0}(c)) and is chosen by a
quantum order-by-disorder mechanism from these classical Fock
states. For $\lambda>\lambda_c$, the ground state breaks
$\mathbb{Z}_2$ symmetry; it consists of Fock states where all
Rydberg excitations are localized in any one of the two rungs of the
ladder (Fig.\ \ref{fig0}(b)). In contrast, for $\lambda <\lambda_c$,
we find a unique ground state which constitutes a macroscopic
superposition of Fock states with fixed $n=1/4$. One such state is
schematically shown in Fig.\ \ref{fig0}(c). This necessitates the
presence of a quantum phase transition which belongs to the 2D Ising
universality class at $\lambda=\lambda_c$. We carry out a
Schrieffer-Wolff transformation to obtain an effective low-energy
Hamiltonian which provides analytical insight into this transition
and identifies the competing terms responsible for it. Our
perturbative analysis, based on the effective Hamiltonian, estimates
the critical point to be $\lambda_c \simeq \Delta/3$ which shows an
excellent match with exact numerical results based on ED. Thus our
analysis indicates that the order-by-disorder mechanism necessarily
leads to a quantum phase transition separating the above-mentioned
ground states. As we detail in the appendix, our study naturally
leads to a class of spin models involving ladders with $\ell_0$ legs
that realize ground states with broken $\mathbb{Z}_{\ell_0}$
symmetry. For $\ell_0=3$, as we show both using ED and analytic
perturbation theory in the appendix, this provides a route to
realization of a quantum critical point belonging to the three-state
Potts universality class in 2D. To the best our knowledge, such
emergent transitions arising from an order-by-disorder mechanism
have not been reported in the context of Rydberg systems in the
literature
\cite{rydexp1,rydexp2,rydexp3,rydexp4,roopayan1,subir2,subir3,other1,ryd2d1,ryd2d2,ryd2d3,ryd2d4,ryd2d5,ryd2d6}.

Next, in an attempt to make contact with realistic experimental
systems, we discuss the fate of this transition for the two-leg
ladder in the presence of a realistic vdW interaction characterized
by $V_{\bf r \bf r'}=V_0/|\bf r -\bf r'|^6$. We discuss an
experimentally relevant regime where $V_0 \gg \lambda, \Delta, w$ so
that states with one or more nearest-neighbor Rydberg excitations do
not feature in the ground state manifold. However, Rydberg
excitations on second and higher neighboring sites now experience a
finite repulsive interaction. We show that the main effect of having
such a second and higher neighbor interaction is to shift
$\lambda_c/\Delta$ to a higher value; the Ising transition still
persists for a wide range of $V_0$ and $w$. We analyze the
transition using a Van Vleck perturbation theory, discuss the
significance of the order-by-disorder mechanism for its stability,
and provide an estimate of $\lambda_c$ as a function of $V_0$ and
$w$. Finally, we discuss experiments which can test our theory.

The plan of the rest of the paper is as follows. In Sec.\ \ref{mod},
we define the model Hamiltonian for Rydberg atoms on a two-leg
ladder. Next, in Sec.\ \ref{modph}, we analyze the phases of the
model by ignoring the second and higher neighboring repulsive
interaction between the Rydberg atoms. This is followed by Sec.\
\ref{pot} where we discuss the effect of finite $V_0$. Finally, we
discuss our main results, point out possible experiments which can
test our theory, and conclude in Sec.\ \ref{diss}. We discuss our
variational wavefunction results around $\lambda \simeq \Delta$ in
App.\ \ref{varwav1} and chart out the phases of ladders with $\ell_0
>2$ rungs, which could be more challenging to realize within
current experimental setups, in App.\ \ref{multileg}.

\section{Model Hamiltonian}
\label{mod}

In this section, we outline the model Hamiltonian for the Rydberg
atoms. We consider an arrangement of two Rydberg chains each having
$2L$ sites shown schematically in Fig.\ \ref{fig0}(a). The sites of
these chains host $N= 4L$ Rydberg atoms whose low-energy effective
Hamiltonian is written as \cite{rydexp1,rydexp2,rydexp3,rydexp4}
\begin{eqnarray}
H &=& \sum_{j=1}^{2L} \sum_{\ell=1}^{2} \left(w \sigma_{j,\ell}^x
-\frac{1}{2}[\Delta (-1)^j + \lambda]\sigma_{j,\ell}^z \right)  +
\sum_{\bf r \ne \bf r'} V_{\bf r \bf r'} \hat n_{\bf r} \hat n_{\bf
r'}, \label{rydham1}
\end{eqnarray}
where $\sigma_{j,\ell}^{\alpha}$ for $\alpha=x,y,z$ denotes the
usual Pauli matrices on the site $j$ of the $\ell^{\rm th}$ chain
and the lattice spacing $a$ between the sites is set to unity so
that the coordinate $\bf r$ of any site is denoted by integers $j$
and $\ell$: ${\bf r}=(j,\ell)$. In Eq.\ \ref{rydham1}, $\lambda$ and
$\Delta$ denote the amplitudes of uniform and staggered detuning
such that $\lambda+\Delta$ and $\lambda-\Delta$ represent the energy
differences between the energy of the applied external laser and the
energy gap between ground and excited Rydberg atomic levels on even
and odd sites respectively. Here $w>0$ denotes the coupling strength
between the Rydberg ground and excited states. In an experimental
setup, this coupling is controlled by two-photon processes having
Rabi frequency $w/\hbar$ \cite{rydexp1,rydexp2,rydexp3,rydexp4}.
Thus for any site with coordinate $\bf r$,
\begin{eqnarray}
\sigma_{\bf r}^x &\equiv& \sigma_{j,\ell}^x = (|G\rangle_{\bf r}\,
_{\bf r}\langle R| +{\rm h.c.}),\quad \hat n_{\bf r} \equiv \hat
n_{j,\ell} = (1+ \sigma_{j,\ell}^z)/2,
\end{eqnarray}
where $|R\rangle_{\bf r} \equiv |\uparrow_{\bf r}\rangle$ and
$|G\rangle_{\bf r} \equiv |\downarrow_{\bf r}\rangle$ denotes
excited and ground states of a Rydberg atom on the site $\bf r$
respectively and $\hat n_{\bf r}$ is the number operator
corresponding to the Rydberg excitations. In the Rydberg excited
state, the atoms experience strong vdW repulsion which is modeled by
$V_{\bf r \bf r'}$ given by
\begin{eqnarray}
V_{\bf r \bf r'} &=& V_0/|\bf r-\bf r'|^6, \label{vdweq}
\end{eqnarray}
where $V_0$ is the interaction strength. It is well-known that in a
typical experiment, $V_0$ can be tuned to induce Rydberg blockade
between neighboring sites. For the rest of this paper, we shall
assume that $V_0 \gg \Delta_0, \lambda, w$ so that the neighboring
sites of the two-rung ladder have at most one Rydberg excited atom
(Fig.\ \ref{fig0}(a)). In this limit, one can split the Hamiltonian
in two parts $H_a$ and $H_b$ given by
\begin{eqnarray}
H_a &=& \sum_{\bf r, \bf r'} V_0 \hat n_{\bf r} \hat n_{\bf r'}
\delta_{|{\bf r}-{\bf r'}|-1},
\nonumber\\
H_b &=& \sum_{j=1}^{2L} \sum_{\ell =1}^{2} \left(w \sigma_{j,\ell}^x
-\frac{1}{2}[\Delta (-1)^j - \lambda]\sigma_{j,\ell}^z \right) +
\frac{V_0}{2} \sum_{\bf r \bf r'} \frac{n_{\bf r} \hat n_{\bf
r'}}{|\bf r-\bf r'|^6} (1-\delta_{|{\bf r}-{\bf r'}|-1}).
\label{splitham1}
\end{eqnarray}
In the limit, where $V_0$ is the largest energy scale in the
problem, it is possible to define a hierarchy of energies for the
eigenstates based on the number of nearest-neighbor Rydberg
excitations \cite{rydexp3}. This can be encoded via a projection
operator $T_m$ which satisfies $[H_a,T_m]= m V_0 T_m$. The role of
$T_m$ is to project the Hamiltonian into a sector of $m$-nearest
neighbor Rydberg excitation. In what follows, we shall be concerned
with the effective Hamiltonian in the sector of $m=0$
nearest-neighbor dipoles. A straightforward calculation similar to
the one carried out in Ref.\ \cite{rydexp3} for a Rydberg chain
yields \cite{subir1a,fss}
\begin{eqnarray}
H_{\rm eff}^V &=& \sum_{j=1}^{2L} \sum_{\ell =1}^{2} w \tilde
\sigma_{j,\ell}^x - \sum_{j=1}^{2L} \sum_{\ell =1}^{2}\frac{1}{2}
[\Delta (-1)^j + \lambda]\sigma_{j,\ell}^z + \frac{V_0}{2} \sum_{\bf
r \bf r'} \frac{n_{\bf r} \hat n_{\bf r'}}{|\bf r-\bf r'|^6}
(1-\delta_{|{\bf r}-{\bf r'}|-1}),\nonumber\\
\tilde \sigma^x_{j,\ell} &=& P_{j-1, \ell} \left(\prod_{\ell'\ne
\ell} P_{j \ell'} \sigma^x_{j,\ell}\right) P_{j+1,\ell},
\label{zeroproj1}
\end{eqnarray}
where we have used the fact that $H_0$ does not contribute in this
sector and $P_{j,\ell}=(1-\sigma^z_{j,\ell})/2$ is the local
projection operator which ensures the absence of nearest-neighbor
Rydberg excitations. The higher-order terms can be systematically
computed involving $m \ne 0$ nearest-neighbor Rydberg excitations
but are unimportant for the regime that we are interested.
Furthermore, in the regime where $V_0 \gg \lambda, \Delta \gg w \gg
V_0/(\sqrt{2})^6$, it is possible to neglect the last term in
$H_{\rm eff}$. In this regime one obtains the following generalized
PXP Hamiltonian\cite{subir1a,fss} on the $2$-leg ladder

\begin{eqnarray}
H_{\rm eff} &=& \sum_{j=1}^{2L} \sum_{\ell =1}^{2} w \tilde
\sigma_{j,\ell}^x - \frac{1}{2} \sum_{j=1}^{2L} \sum_{\ell
=1}^{2}[\Delta (-1)^j + \lambda]\sigma_{j,\ell}^z.  \label{effham1}
\end{eqnarray}
In the next section, we shall obtain the ground state phase diagram
of $H_{\rm eff}$ given in Eq.\ \ref{effham1}. The effect of finite
second and higher-neighboring interaction shall be discussed, using
$H_{\rm eff}^V$ (Eq.\ \ref{zeroproj1}) in Sec.\ \ref{pot}.

\section{Phases of $H_{\rm eff}$}
\label{modph}

To understand the phase-diagram of $H_{\rm eff}$, we first set
$w=0$. It is then easy to see that for $\lambda<0$ and $|\lambda|
\gg \Delta$, we have $\langle \sigma_{j,\ell}^z \rangle =-1$ on
all sites leading to $n=\langle \sum_{\vec r} \hat n_{\vec
  r}\rangle/N= 0$. Similarly for $\lambda>0$ and $\lambda \gg \Delta$,
the ground states hosts maximal number of possible up-spins or
Rydberg excitations. However, due to the constraint of having at
most one Rydberg excitation on neighboring sites, it can have such
excitations only on $N/2$ sites. This results in a net Rydberg
excitation density of $n= 1/2$ and leads to a two-fold degenerate
ground state. In between for $-\Delta \le \lambda \le \Delta$, the
ground state hosts one Rydberg excitation on every even rung of the
ladder and thus has $n=1/4$. This is due to the fact that such
excitations requires $\delta E_{\rm even}= -(\lambda+\Delta) <0$ in
this regime. In contrast, a Rydberg excitation on an odd site costs
$\delta E_{\rm odd} = \Delta-\lambda>0$. The ground state manifold
therefore has macroscopic $2^{L}$ fold classical degeneracy for
$w=0$. At $\lambda=\pm \Delta$, $n$ exhibits jumps for $w=0$ which
constitute first order transitions.

\begin{figure}[ht]
\centering
\includegraphics[width=0.49\linewidth]{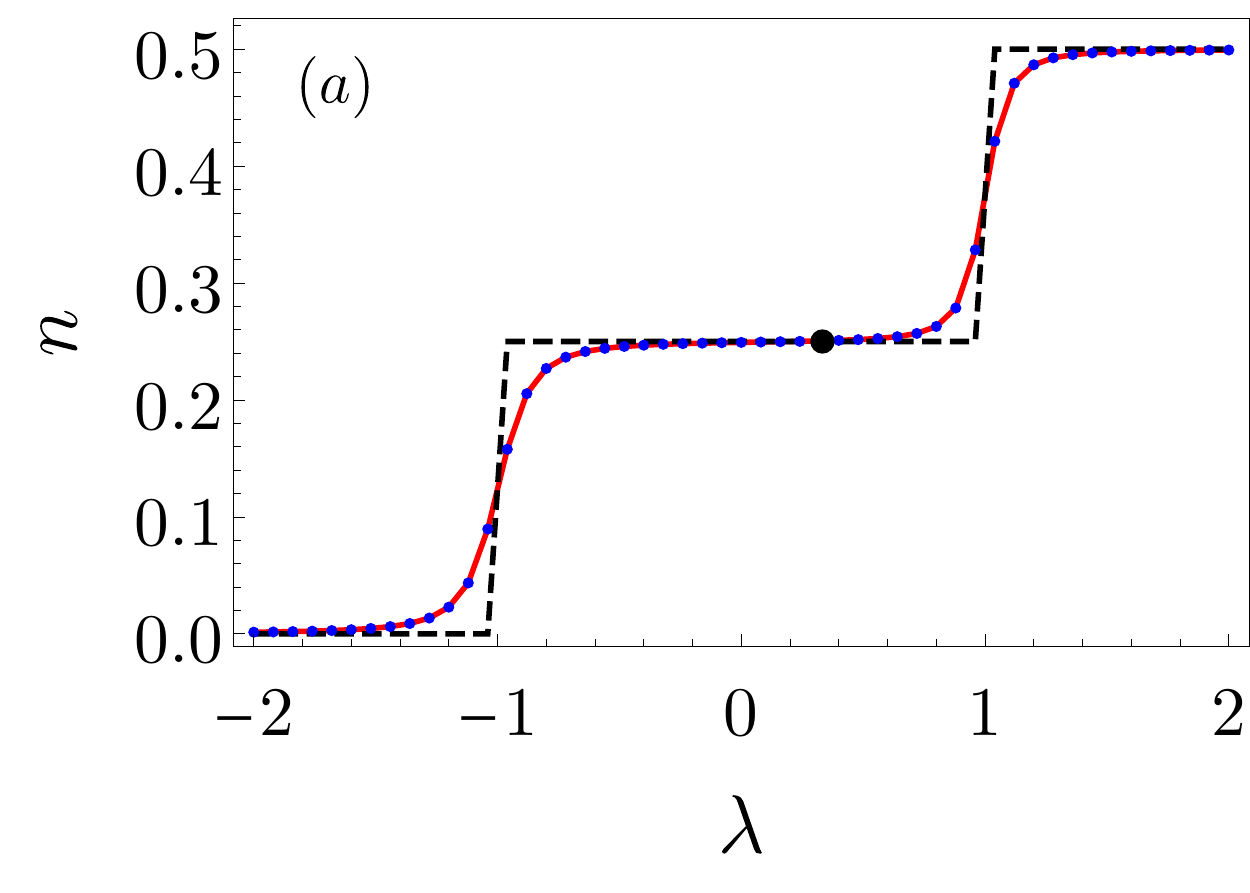}
\includegraphics[width=0.49\linewidth]{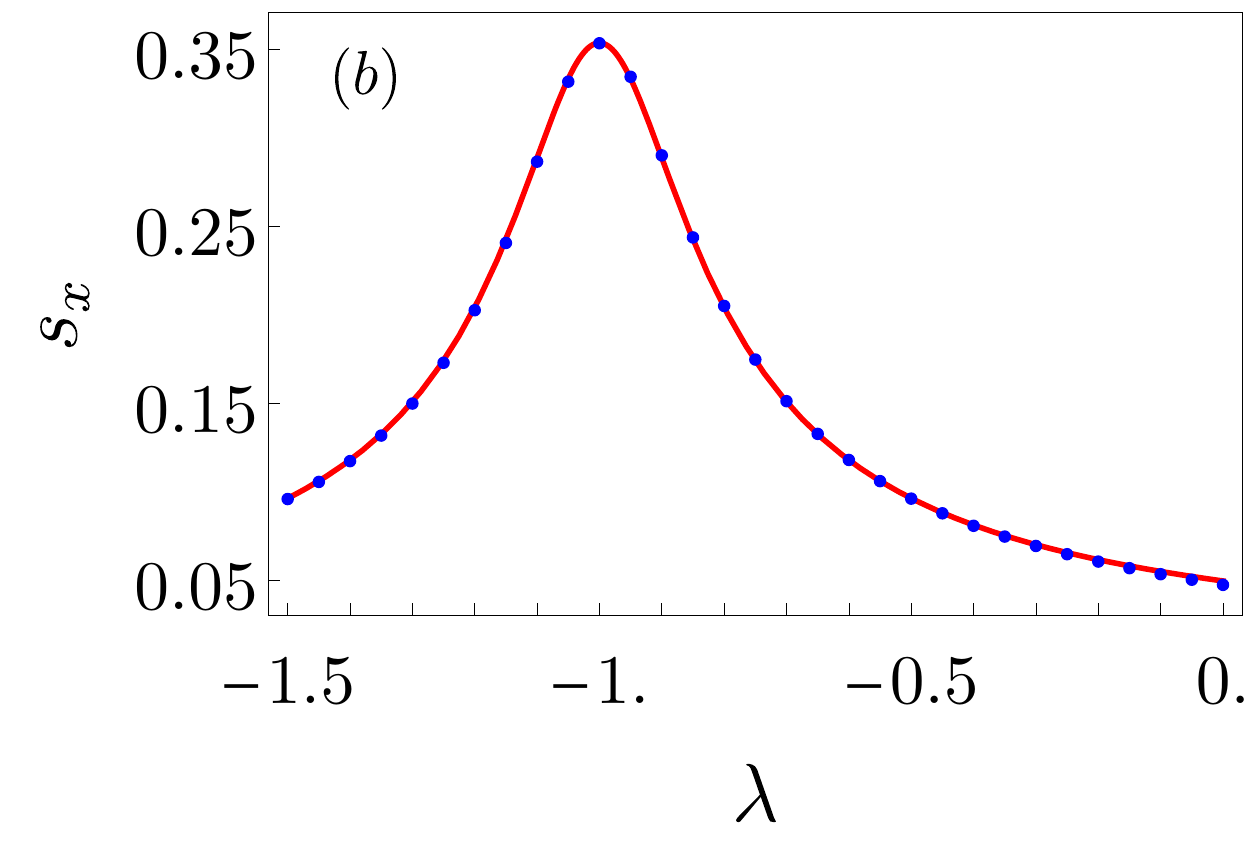}
\caption{(a): Plot of the Rydberg excitation density $n$ as a
function of $\lambda$ with $w=0$ (black dotted line) and $w=0.05$
(blue dots) for the two leg-ladder indicating the plateau at with
$n=1/4$ between $-1\le \lambda \le 1$ as obtained using ED. The red
solid line indicates $n$ as computed using variational wavefunction
analysis around $\lambda= \pm 1$.  The black circle at $\lambda \sim
0.33$ shows the position of the Ising transition. (b): Plot of $s_x$
as a function of $\lambda$ across the crossover at $\lambda \simeq
-1$. The solid red line indicate results from the variational
wavefunction approach while the blue dots indicates ED results
corresponding to  $N=20$. All energies are scaled in units of
$\Delta$. See text for details.} \label{fig1}
\end{figure}

These discontinuous transitions become smooth crossovers due to
quantum fluctuations introduced by a finite $w$ (Fig.~\ref{fig1}
(a)). This can be understood using a variational wavefunction based
analysis. Here we illustrate this for $\lambda/\Delta \simeq -1$; a
similar analysis can be carried out for $\lambda \simeq \Delta$ and
is shown in App.\ \ref{varwav1}. We start by noting that for
$|\lambda|\gg \Delta,w$ and $\lambda <0$, the ground state of
$H_{\rm eff}$ is given by
\begin{eqnarray}
|\psi_1\rangle = \prod_{j=1}^{L} |\downarrow_{2j-1,1}
\downarrow_{2j-1,2}; \downarrow_{2j,1} \downarrow_{2j,2} \rangle.
\label{vrw1}
\end{eqnarray}
In contrast, for $\Delta>|\lambda|$, it is clearly energetically
favorable to have a Rydberg excitation on even sites. A variational
wavefunction representing such a state is given by
\begin{eqnarray}
|\psi_2\rangle &=& \prod_{j=1}^{L} \left(\cos \phi
|\downarrow_{2j-1,1} \downarrow_{2j-1,2}; \uparrow_{2j,1}
\downarrow_{2j,2} \rangle + \sin \phi |\downarrow_{2j-1,1}
\downarrow_{2j-1,2}; \downarrow_{2j,1} \uparrow_{2j,2}
\rangle\right). \label{state2}
\end{eqnarray}
Note that we have chosen the variational parameter $\phi$ to be
independent of position since we intend to carry out a mean-field
analysis of the problem here. Near $\lambda/\Delta=-1$, we construct
a variational wavefunction $|\psi_v\rangle = \cos \theta
|\psi_1\rangle + \sin \theta |\psi_2(\phi)\rangle$ which leads to
$E_v =\langle \psi_v|H |\psi_v\rangle$ given by
\begin{eqnarray}
E_v &=&  (\lambda \cos^2 \theta -  \Delta \sin^2 \theta) - \sqrt{2}w
\sin 2\theta \cos (\phi-5\pi/4), \label{varen}
\end{eqnarray}
where we have ignored an irrelevant constant term. The minimization
of $E_v$ fixes
\begin{eqnarray}
\phi &=& \phi_0=5\pi/4, \quad \theta=\theta_0=\frac{1}{2} \arctan
\left[\frac{-2 \sqrt{2} w}{\lambda+\Delta}\right]. \label{varpar}
\end{eqnarray}

To show the efficacy of this approach, we compare results obtained
via variational wavefunction method with exact numerics. To this
end, we analyze $H_{\rm eff}$ numerically using ED for $N\le 40$ and
using periodic boundary condition along the chains. We compute the
expectation values $\hat n$ and
\begin{eqnarray}
\hat s_x= \sum_{j=1}^{L} \sum_{\ell =1}^2\sigma_{2j,\ell}^{x}/N,
\end{eqnarray}
from both the variational wavefunction and using exact numerics. The
former yields analytical expressions for $n$ and $s_x =\langle \hat
s_x \rangle $ given by
\begin{eqnarray}
n &=& \frac{1}{8} \sin^2 \theta_0, \quad s_x = \frac{ \sqrt{2}\sin 2
\theta_0}{4}. \label{expec}
\end{eqnarray}
The plots of these as a function of $\lambda/\Delta$ with
$w/\Delta=0.05$ are compared with their exact numerical counterparts
in Fig.\ \ref{fig1}(a) and \ref{fig1}(b). The plots show an
excellent match; this shows that the crossover between phases with
$n=0$ to $n=1/4$ is well captured by our variational analysis. A
similar match is obtained for the crossover at $\lambda/\Delta
\simeq 1$ as can be seen from Fig.\ \ref{fig1}(a); this has been
presented in detail in App.\ \ref{varwav1}. We also note here that
the extent of the plateau reduces with increasing $w$; we therefore
work in the regime $w \ll \lambda, \Delta$ in this work.

\begin{figure}[ht]
\centering
\includegraphics[width=0.49\linewidth]{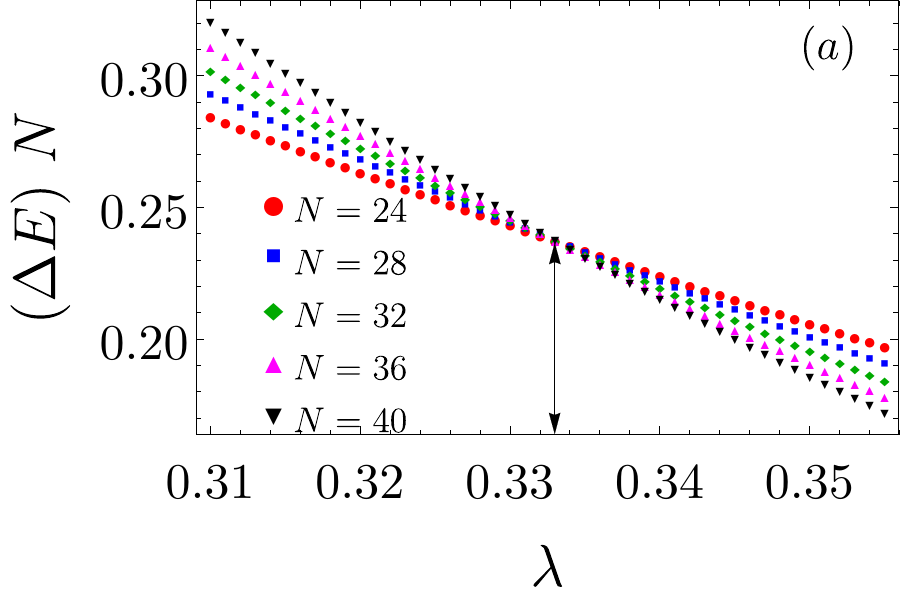}
\includegraphics[width=0.49\linewidth]{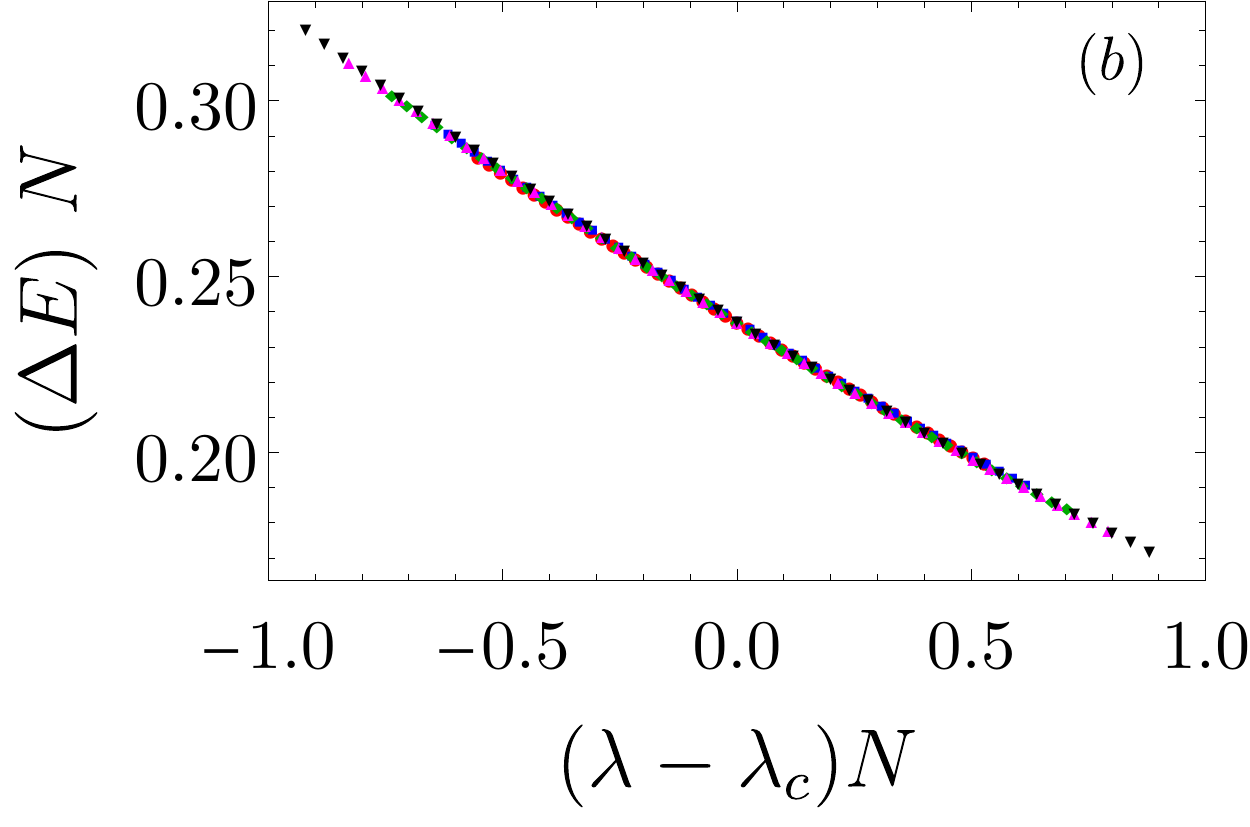}
\includegraphics[width=0.49\linewidth]{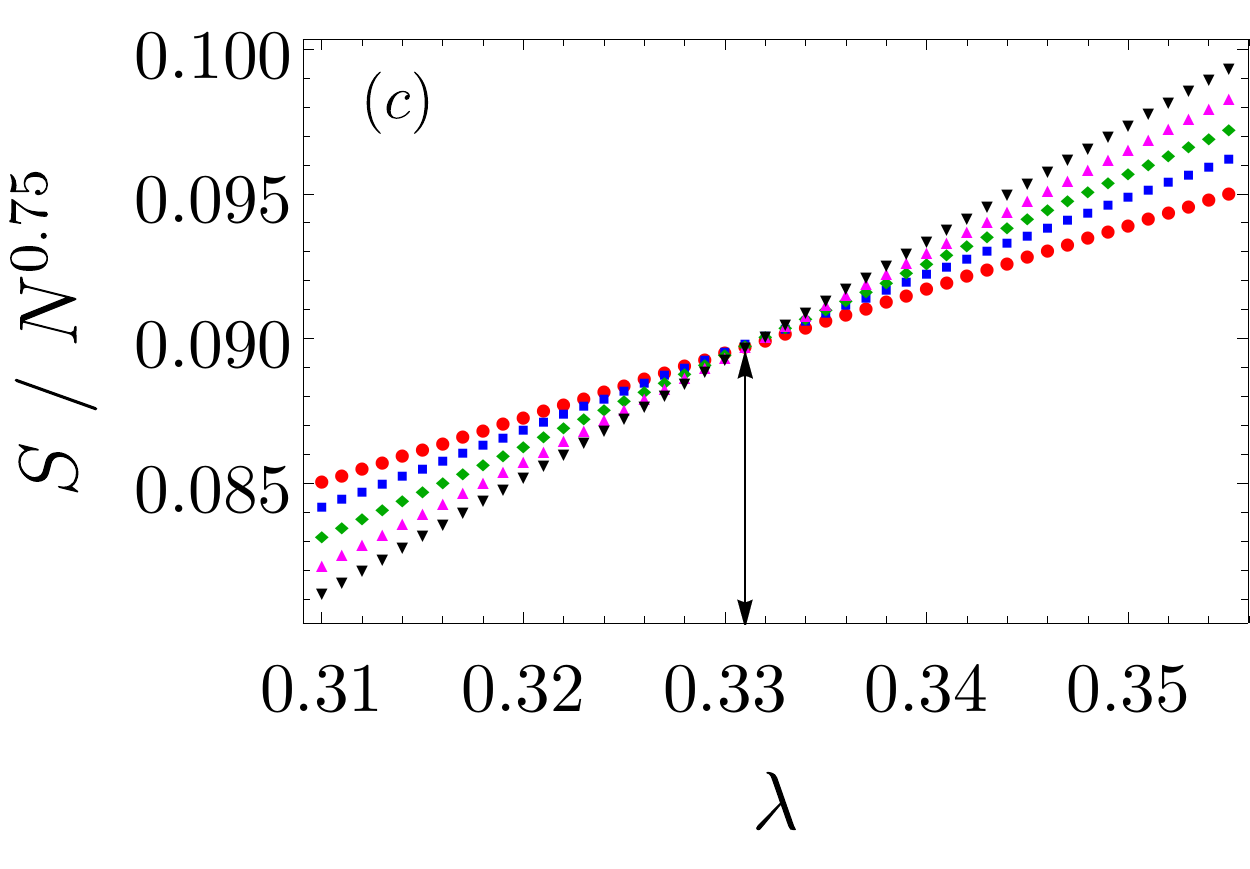}
\includegraphics[width=0.49\linewidth]{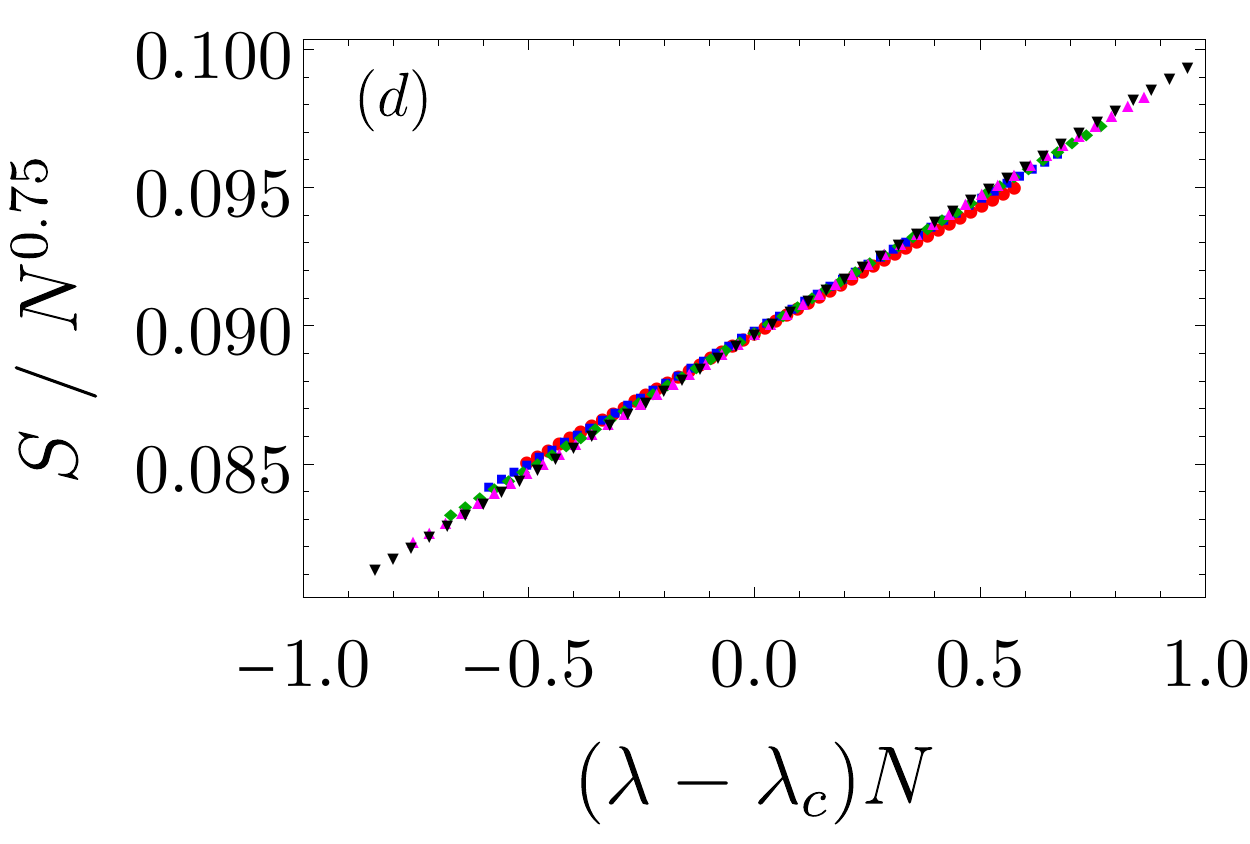}
\caption{(a): Plot of $(\Delta E) N^z$ for a two-leg ladder as a
function of $\lambda$ for $w=0.05$ for several $N$ showing a
crossing at $\lambda=\lambda_c \simeq 0.333$ for $z=1$. (b) Plot of
$(\Delta E) N^z$ as a function of $N^{1/\nu}(\lambda-\lambda_c)$
showing perfect scaling collapse for $z=\nu=1$. (c): Plot of $S
N^{2-z-\eta}$ as a function of $\lambda$ showing a crossing at
$\lambda_c \simeq 0.331$ for $\eta=0.25$. (d) Plot of $S
N^{2-z-\eta}$ as a function of $(\lambda-\lambda_c)N^{1/\nu}$
showing perfect scaling collapse for $\nu=1$ and $\eta=0.25$.  All
energies are scaled in units of $\Delta$. See text for details.}
\label{fig2}
\end{figure}

Next, we concentrate on the nature of the ground state of the system
for $-1\le \lambda/\Delta \le 1$. Here the ground state is chosen by
a quantum order-by-disorder mechanism from the manifold of $2^L$
classical Fock states with $n=1/4$. Our numerical analysis reveals
the presence of a quantum phase transition inside the plateau with
$n=1/4$ at $\lambda=\lambda_c$. We find numerically that the ground
state of the system at $\lambda>\lambda_c$ hosts Rydberg excitations
on even sites of {\it any one of the two chains}; it thus breaks a
$\mathbb{Z}_2$ symmetry as shown schematically in Fig.\
\ref{fig0}(b). In contrast for $\lambda < \lambda_c$, the ground
state is unique and constitute a superposition of all Fock states
with $n=1/4$ (Fig.\ \ref{fig0}(c)); it does not break any symmetry.
This necessitates a transition at $\lambda=\lambda_c$ characterized
by an order parameter
\begin{eqnarray}
\hat O &=& \sum_{j=1}^{L} \sum_{\ell =1}^{2} (-1)^{\ell}
\sigma_{2j,\ell}^z. \label{opeq}
\end{eqnarray}
We note that the quantum phase transition is a result of the fact
that the order-by disorder mechanism chooses two different ordered
states, one of which breaks a discrete $\mathbb{Z}_2$ symmetry while
the other does not.

%%%%%%%%%%%%%%%%%%%%%%%%%%%%%%%%%%%%%%%%%%%%%%%%%%%%%%%%%%
\begin{figure}[ht]
\centering
\includegraphics[width=\linewidth]{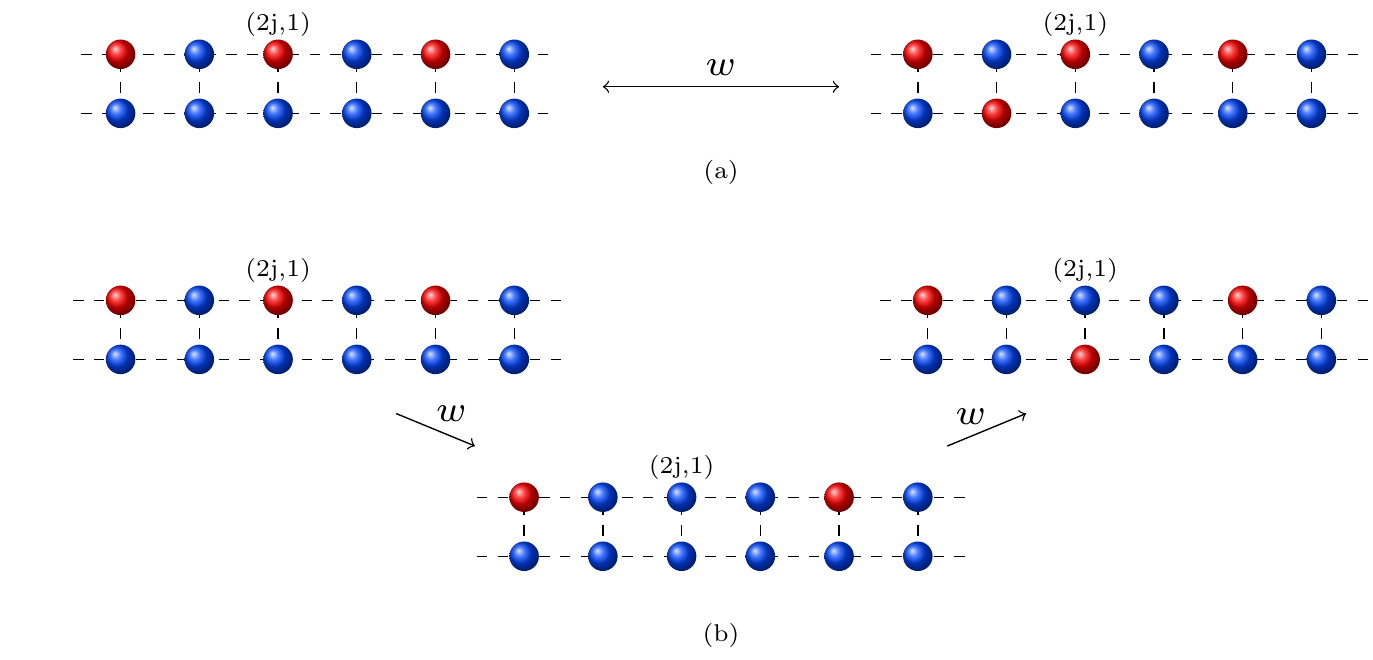}
\caption{ Schematic representation of the virtual processes leading
to the effective Hamiltonian $H'$. The red (blue) circles represent
atoms in Rydberg excited (ground) states. (a): The process which
favors all Rydberg excitations to be on the even sites of one of the
chains (chosen to be the top chain ($\ell=1$) here) and leads to the
first term in $H'$. The excited state constitutes one additional
Rydberg excitation on an odd site of the bottom chain as shown
schematically. (b) The competing virtual process which leads to the
second term of $H'$. The excited state, which belongs to the high
energy manifold, constitutes one additional Rydberg atom in its
ground state at an even site of the top chain as shown. See text for
details.} \label{fig4}
\end{figure}
%%%%%%%%%%%%%%%%%%%%%%%%%%%%%%%%%%%%%%%%%%%%%%%%%%%%%%%%%%%%%

To check for the universality class of this transition, we carry out
finite-size scaling analysis for $N\le 40$ on these ladders. We use
the well-known scaling relations \cite{scalingref,subir1a}
\begin{eqnarray}
\Delta E &=& N^{-z} f\left[N^{1/\nu}(\lambda-\lambda_c)\right],
\label{scalingeq}\\
S &=& \frac{1}{N} \langle \hat O^2 \rangle = N^{2-z-\eta}
g\left[N^{1/\nu}(\lambda-\lambda_c)\right], \nonumber
\end{eqnarray}
where $z$ and $\nu$ are the dynamical critical and the correlation
length exponents respectively, $\eta$ indicates the anomalous
dimension of $\hat O$, $S$ is the order-parameter correlation, $f$
and $g$ are scaling functions, and $\Delta E$ denotes the energy gap
between the ground and the first excited states. Such a scaling
analysis, shown in Fig.\ \ref{fig2} indicates the presence of a
quantum phase transition at $\lambda_c \simeq 0.33 \Delta$ with
$z=\nu=1$ and $\eta=1/4$; these exponents confirm that the
universality class is the same as the critical point of the
classical 2D Ising model.

To obtain analytic insight into this transition, we construct a
perturbative  effective Hamiltonian using a Schrieffer-Wolff
transformation as follows. First, we note that $w\ll |\lambda|,
\Delta$ in the regime of interest. Using Eq.\ \ref{effham1}, we
therefore write $H_{\rm eff} =H_0 +H_1$ where $H_1= w$
$\sum_{j,=1}^{2L} \sum_{\ell =1}^{2} \tilde \sigma_{j,\ell}^x$. We
then use a canonical transformation operator $S'$ to write
\begin{eqnarray}
H' =\exp[iS'] H_{\rm eff} \exp[-iS']= H_0 +H_1 +[iS',H_0+H_1]
+\frac{1}{2}[iS',[iS',H_0]] + ...,
\end{eqnarray}
where the ellipsis indicates higher order terms. Next, following
standard procedure, we eliminate all ${\rm O}(w)$ terms in $H'$
which takes one out of the low-energy manifold of states. This
allows to determine $S'$ using the resultant condition $[iS',H_0]=
-H_1$ yielding
\begin{eqnarray}
S'= -w \sum_{j=1}^{2L} \sum_{\ell =1}^{2} (\Delta (-1)^j +
\lambda)^{-1} \tilde \sigma_{j,\ell}^y. \label{caneq}
\end{eqnarray}
Using the expression of $S'$, a straightforward calculation yields
the effective Hamiltonian $H'=H_0 $  $+[iS',[iS',H_0]]/2$. A
subsequent projection to the manifold of states with $n=1/4$ leads
to \cite{conf1,ksen1}
\begin{eqnarray}
H' &=& \frac{-w^2}{\Delta-\lambda} \sum_{j=1}^{L}  \Big[ \sum_{\ell
=1}^{2} P_{2j, \ell} P_{2j+2,\ell} + \frac{\alpha_0}{2} \sum_{\ell
,\ell'=1,2; \ell \ne \ell'} P_{2j, \ell'} (\sigma^x_{2j, \ell}
\sigma^x_{2j, \ell'} +\sigma^y_{2j, \ell} \sigma^y_{2j, \ell'}
)\Big], \label{effproj}
\end{eqnarray}
where $\alpha_0= (\Delta-\lambda)/(\Delta+\lambda)$. The virtual
processes induced by $w$ which play a key role for realization of
these two terms are schematically shown in Fig.\ \ref{fig4}.

The emergence of the phase transition at a critical value of
$\lambda/\Delta$ can be understood by noting the competition between
the two terms of $H'$ in Eq.\ \ref{effproj}. The first term, whose
amplitude dominates for $\Delta -\lambda \simeq w$, prefers maximum
possible Rydberg excitation on even sites of any one of the two
chains; such a state is schematically shown in Fig.\ \ref{fig0}(b).
The resultant ground state thus breaks $\mathbb{Z}_2$ symmetry. In
contrast, the second term, which dominates when $\Delta +\lambda
\simeq w$, prefers a linear superposition of all states with one
Rydberg excitation on any chain of even sites of the ladders. The
resultant ground state does not break any discrete symmetry. This
necessitates an intermediate quantum critical point belonging to the
2D Ising universality class.

\begin{figure}[ht]
\centering
\includegraphics[width=0.49\linewidth]{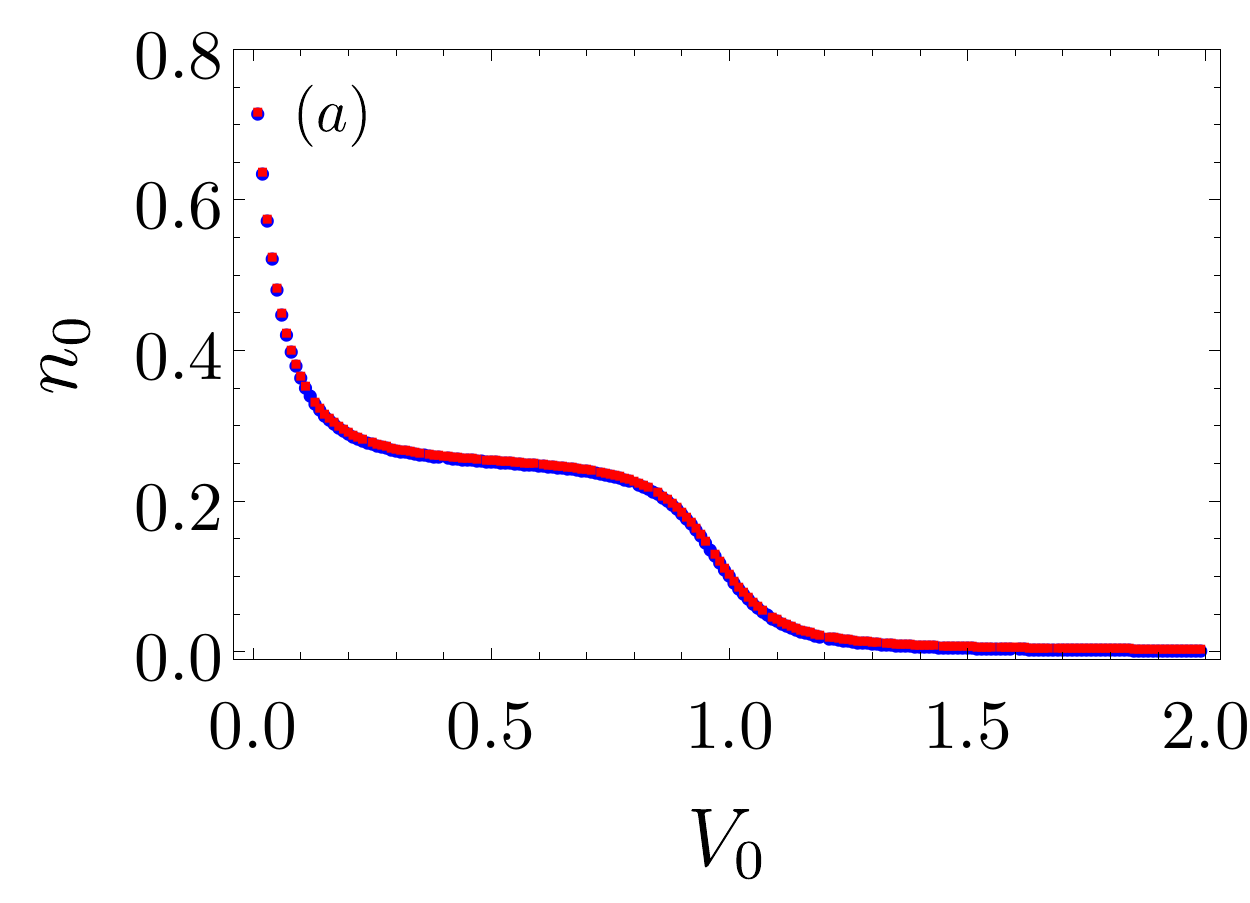}
\includegraphics[width=0.49\linewidth]{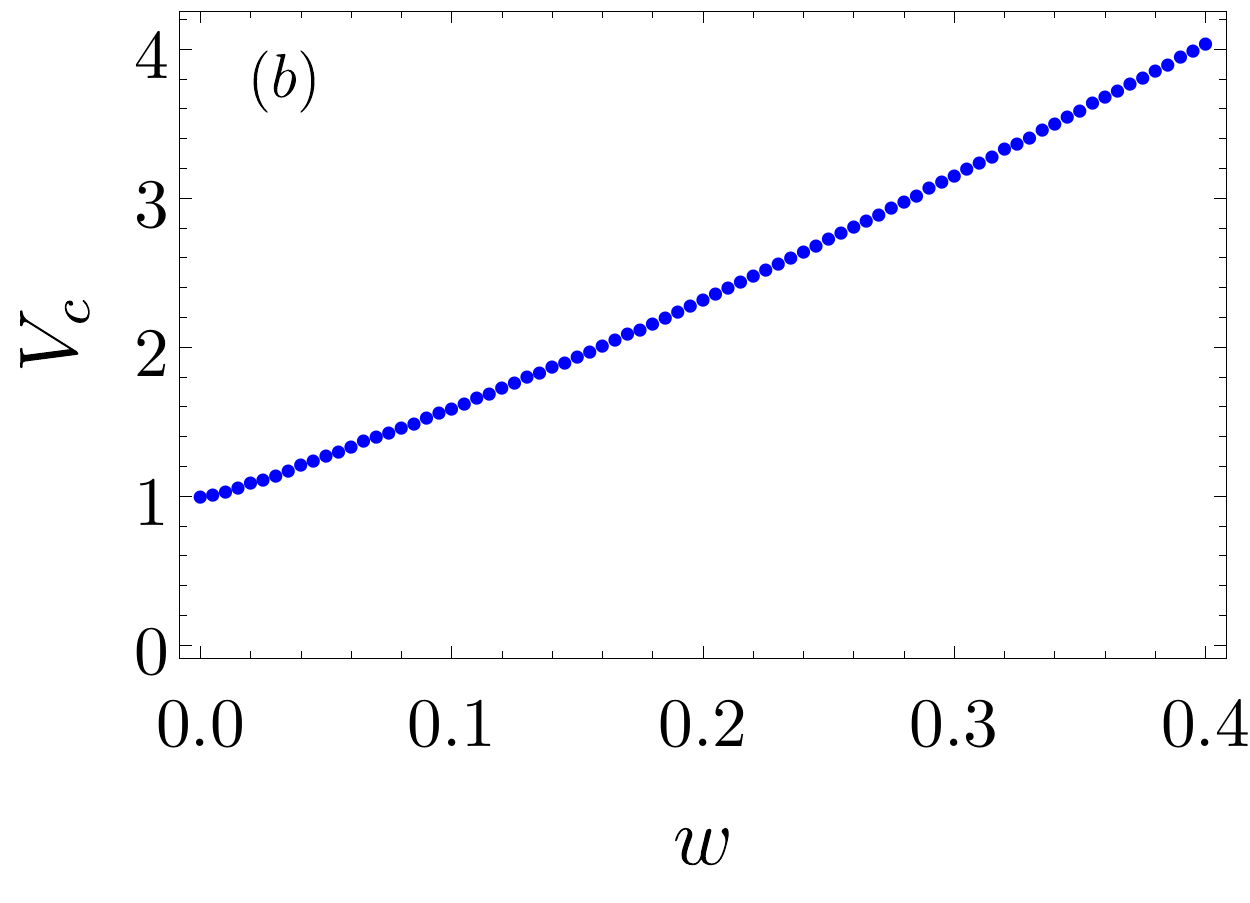}
\includegraphics[width=0.49\linewidth]{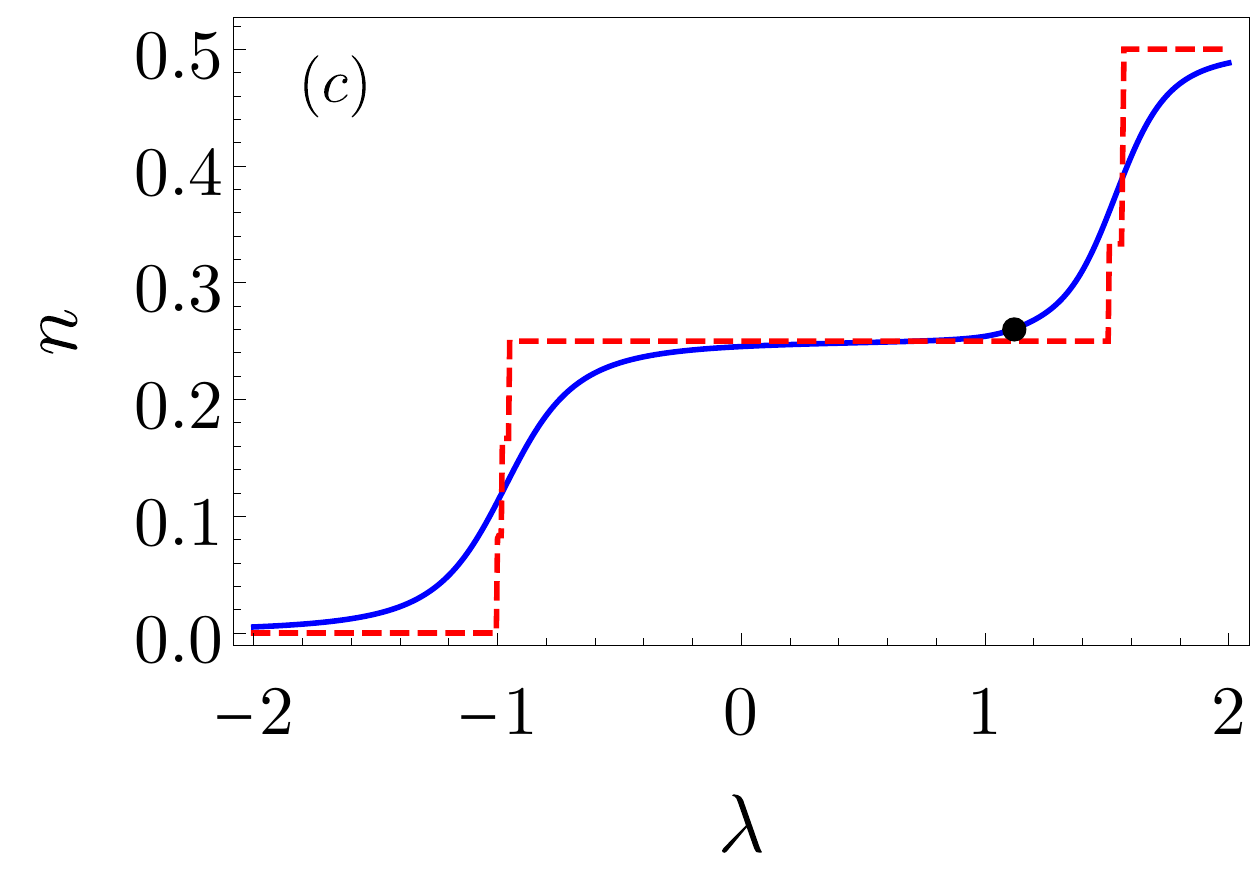}
\includegraphics[width=0.49\linewidth]{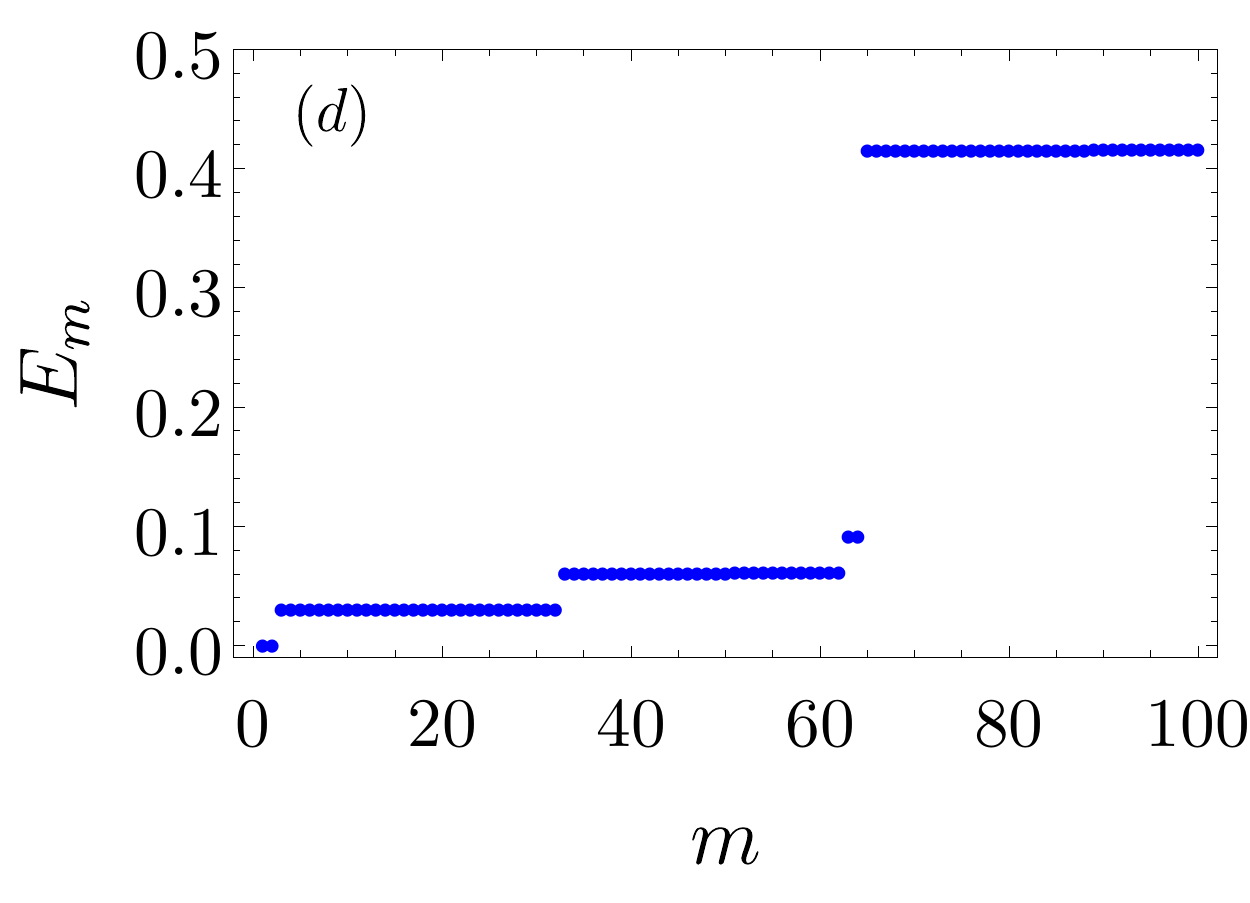}
\caption{(a): Plot of $n_0$ as a function of $V_0$ for $w=0.1$ and
$\lambda=1$ showing $V_c \simeq 1.6$ for $N=12$ (blue squares) and
$N=16$ (red circles). (b) Plot of $V_c$ as a function of $w$ for
$\lambda=1$. (c) Plot of $n$ as a function of $\lambda$ for $w=0.1$
and $V_0=2$ showing the plateau for $n$ and the position of the
transition (black circle).(d) The eigenenergies $E_m$ of $H^V_{\rm
eff}$, measured from the ground state energy, as a function of $m$
(for $m \le 100$) for $\lambda =1$, $w=0$ and $N=24$ showing the
near-degenerate manifold. The two degenerate ground states at $w=0$
correspond to two Fock states where the Rydberg excitations occur
alternately on even sites of ladders $1$ and $2$. All energies are
scaled in units of $\Delta$. See text for details.} \label{fig5}
\end{figure}

To estimate the position of this critical point, we start from the
$\mathbb{Z}_2$ symmetry broken ground state with all Rydberg excitations on
even sites of the first chain (Fig.\ \ref{fig0}(b)). The simplest
excited state over these ground states constitutes equal
superposition of Fock states with one Rydberg excitation on any one
of two chains for a single even site $2j$; for $j' \ne 2j$, the
Rydberg excitations occur on the first chain. This excited state can
be represented as
\begin{eqnarray}
|\psi_{\rm ex}\rangle &=& \frac{1}{\sqrt{2}} \Big( |\downarrow_{1,1}
\downarrow_{1,2}; \uparrow_{2,1} \downarrow_{2,2}
....\uparrow_{2j,1} \downarrow_{2j,2}...;\uparrow_{2L,1}
\downarrow_{2L,2}\rangle \nonumber\\
&& + |\downarrow_{1,1} \downarrow_{1,2}; \uparrow_{2,1}
\downarrow_{2,2} ....\downarrow_{2j,1}
\uparrow_{2j,2}...;\uparrow_{2L,1} \downarrow_{2L,2}\rangle \Big),
\label{exwav}
\end{eqnarray}
where $|\psi_G\rangle =|\downarrow_{1,1} \downarrow_{1,2};
\uparrow_{2,1} \downarrow_{2,2} ....\uparrow_{2j,1}
\downarrow_{2j,2}...;\uparrow_{2L,1} \downarrow_{2L,2}\rangle$ is
the symmetry broken ground state.

The energy cost of creating such an excited state can be easily
computed as $\delta E_{\rm ex} = $ \\ $\langle \psi_{\rm ex}|
H'|\psi_{\rm ex}\rangle -\langle \psi_G|H'|\psi_G\rangle$. A
straightforward calculation yields $\delta E_{\rm ex}= \delta E_1
+\delta E_2$ where
\begin{eqnarray}
\delta E_1 &=& \frac{w^2}{2(\Delta -\lambda)}, \quad \delta E_2 =
-\frac{w^2}{\Delta+\lambda}.\label{encal}
\end{eqnarray}
We note that $\delta E_1$ originates from the first term of $H'$
which prefers all the Rydberg excitations to be on the same chain.
In contrast, $\delta E_2$ comes from the second term which maximizes
superposition of up-spins on different sites of the even rungs. An
approximate estimate of the critical point can be obtained by
equating these two energies; this yields the value of $\lambda$ at
which $\delta E_{\rm ex}=0$ so that the $\mathbb{Z}_2$ symmetry broken ground
state becomes unstable to low-energy excitations. Such an estimate
yields
\begin{eqnarray}
\lambda_c = \Delta/3. \label{crites}
\end{eqnarray}
We note that this simple estimate matches remarkably well with that
obtained from ED using finite-sized scaling analysis of the energy
gap: $\lambda_c^{\rm ex}/\Delta= 0.332$. This provides support for
the fact that the transition is indeed realized due to an
order-by-disorder mechanism which can be captured by the
perturbative effective Hamiltonian $H'$ (Eq.\ \ref{effproj}).

We also note that the position of the transition can also be
obtained as follows. First, we define pseudospin states on even
sites of the ladder as $|1_{2j}\rangle \equiv |\uparrow_{2j,1};
\downarrow_{2j,2}\rangle$ and $|-1_{2j}\rangle \equiv
|\downarrow_{2j,1}; \uparrow_{2j,2}\rangle$. In terms of Pauli
matrices, $\vec s_j$, which acts on the space of these pseudospins,
one can then identify  $P_{2j,1(2)} = (1-(+)s_{2j}^z)/2$. Using this
identification, the first term of $H_{\rm eff}$ (Eq.\
\ref{effproj2}) can be written, ignoring an irrelevant constant
term, as
\begin{eqnarray}
\sum_{j=1}^{L}  \sum_{\ell =1}^{2} P_{2j, \ell} P_{2j+2,\ell} &\to&
\frac{1}{2} \sum_{j=1}^{L} s_{2j}^z s_{2j+2}^z \label{term1}
\end{eqnarray}
We then note that the second term of $H_{\rm eff}$ (Eq.\
\ref{effproj2}), acting on these pseudospin state, yields
$|1_{2j}\rangle \Leftrightarrow |-1_{2j}\rangle$. This is due to the
fact that the spins even sites of the ladder can not be both spin-up
or both spin-down; the first condition follows from the constraint,
while the second is a consequence of the ground state belonging to
$n=1/4$ sector. Using this, one can show
\begin{eqnarray}
\frac{1}{2} \sum_{j=1}^L \sum_{\ell ,\ell'=1,2; \ell \ne \ell'}
P_{2j, \ell'} (\sigma^x_{2j, \ell} \sigma^x_{2j, \ell'}
+\sigma^y_{2j, \ell} \sigma^y_{2j, \ell'} )  &\to&  \sum_{j=1}^L
s_{2j}^x \label{term2}
\end{eqnarray}
Using Eqs.\ \ref{term1} and \ref{term2}, we find an effective Ising
representation of $H_{\rm eff}$ given by
\begin{eqnarray}
H_{\rm eff}^{I} &=& \frac{-w^2}{\Delta-\lambda} \sum_{j=1}^{L}
\left( \frac{1}{2} s_{2j}^z s_{2j+2}^z + \alpha_0 s^x_{2j} \right)
\label{isingeff}
\end{eqnarray}
From this, one can, using Kramers-Wannier duality, read off the
position of the critical point to be at $\alpha_0=1/2$ which also
leads to $\lambda_c= \Delta/3$.

Before ending this section, we would like to point out that a
similar model ladders with $\ell_0>2$ would lead to spin models
which hosts $\mathbb{Z}_{\ell_0}$ symmetry broken ground states
leading to realization of non-Ising quantum critical points. We show
in the appendix, by an analysis similar to that carried out above,
that such transition belong to 2D three-state Potts universality
class for $\ell_0=3$. However, realization of $\ell_0 > 2$-leg
ladders using Rydberg atoms chains with vdW interactions whose
low-energy behavior is governed by $H_{\rm eff}$ may be difficult;
we discuss this further in the appendix.

\section{Effect of vdW interaction}
\label{pot}

In this section, we shall discuss the fate of the Ising transition
for a realistic vdW interaction given by Eq.\ \ref{vdweq}. To this
end, we concentrate in the regime where $V_0$ is large enough so
that there is at most one Rydberg excitation in any pair of
neighboring sites. To estimate the minimal $V_0\equiv V_{c}$
required to satisfy this criteria, we diagonalize $H$ (Eq.\
\ref{rydham1}) for a two-leg ladder with $N \le 16$ and Hilbert
space dimension ${\mathcal D} \le 2^{16}$ to obtain its ground state
$|\Psi_G\rangle$. We then compute
\begin{eqnarray}
n_{0} &=& \frac{1}{2} \sum_{\bf r \bf r'} \langle \Psi_G |n_{\bf r}
n_{\bf r'} \delta_{|{\bf r} -{\bf r'}|-1} |\Psi_G\rangle
\label{n12eq}
\end{eqnarray}
as a function of $V_0$ for a fixed $w$, $\lambda$, and $\Delta$.
This allows us to estimate $V_c(w/\Delta,\lambda/\Delta)$ as the
minimum value of $V_0$ for which $n_0 \le \epsilon_0\simeq 0.001$.
This is shown in Fig.\ \ref{fig5}(a) where $n_0$ is plotted as a
function of $V$ for $\lambda/\Delta =1$ and $w/\Delta=0.1$ for
$N=12,16$. We note that $V_c$ is independent of $N$; thus one can
safely use $V_0 >V_c$ to access the constrained Hilbert space for
larger system sizes. A plot of $V_c/\Delta$ obtained from this
procedure for $N=16$ as a function of $w/\Delta$ for
$\lambda/\Delta=1$ is shown in Fig.\ \ref{fig5}(b). We find that
$V_c$ increases with increasing $w$; this can be easily understood
from the fact that increasing $w$ makes the freezing of spin on any
site energetically more costly.

Next, we fix $V_0/\Delta=2$ and $w/\Delta=0.1$ so that the ground
state does not have more than one Rydberg excitation on neighboring
sites. We then diagonalize $H_{\rm eff}^V$ (Eq.\ \ref{zeroproj1})
for $N=40$, obtain the ground state, and compute $n$ as a function
of $\lambda/\Delta$ for $w=0$ and $0.1 \Delta$. Our results, shown
in Fig.\ \ref{fig5}(c), indicate the presence of a wide range of
$\lambda/\Delta$ with $n=1/4$; for $w/\Delta=0.1$, $n$ starts
deviating from $1/4$ around $\lambda/\Delta \simeq 1.1$ and reaches
its maximal value $1/2$ around $\lambda_u/\Delta \sim 2$. Our
numerical, finite-size scaling analysis, also finds a Ising critical
point around $\lambda=\lambda_c$ such that $\lambda_c/\Delta \simeq
1.12$ at the edge of the plateau. For $\lambda_c <\lambda <
\lambda_u$, the system has a double degenerate ground state; it
chose a single configuration only via spontaneous symmetry breaking.
This shows that the presence a finite second and higher neighbor
interaction does not necessarily obliterate the Ising transition;
however the transition is shifted to a higher value of
$\lambda/\Delta$. We note that this is possible since the residual
interaction term in $H_{\rm eff}^V$ does not lift the degeneracy
between the two ${\mathbb Z}_2$ symmetry broken ground states.
Importantly, the presence of $V$ leads to a near-degenerate
low-energy manifold with ${\rm O}(L)$ states having $n \sim 1/4$;
this is to be contrasted with the exact degenerate ground state
manifold of $2^L$ states when the next nearest-neighbor interactions
are neglected. This is shown in Fig.\ \ref{fig5}(d) for
$\lambda/\Delta = 1.1$ and $w=0$. The ground states for $w=0$ shown
in Fig.\ \ref{fig5}(d) correspond to two Fock states where Rydberg
excitations occur alternatively on even sites of ladders $1$ and
$2$. We also note that the presence of a finite $w$, whose magnitude
is comparable to the width of the nearly-degenerate low-energy
manifold, leads to a different ground state which constitutes a
significant admixture of states within this manifold. The plot also
indicates that one needs a finite but small $w$ so that states with
$n\ne 1/4$ (higher excited states in Fig.\ \ref{fig5}(d)) are not
strongly mixed with near degenerate  ${\rm O}(L)$ manifold of
low-lying states.

\begin{figure}[ht]
\centering
\includegraphics[width=0.49\linewidth]{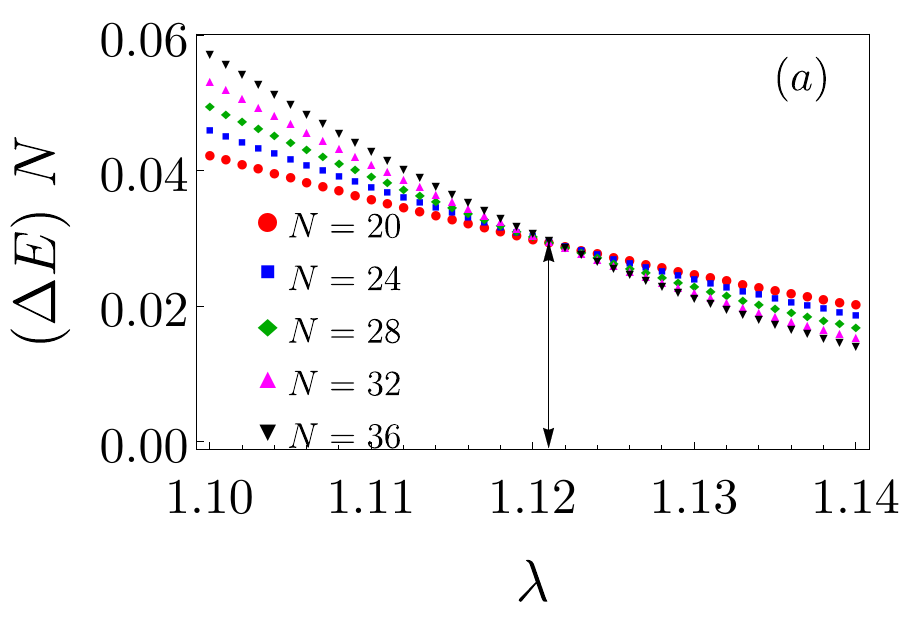}
\includegraphics[width=0.49\linewidth]{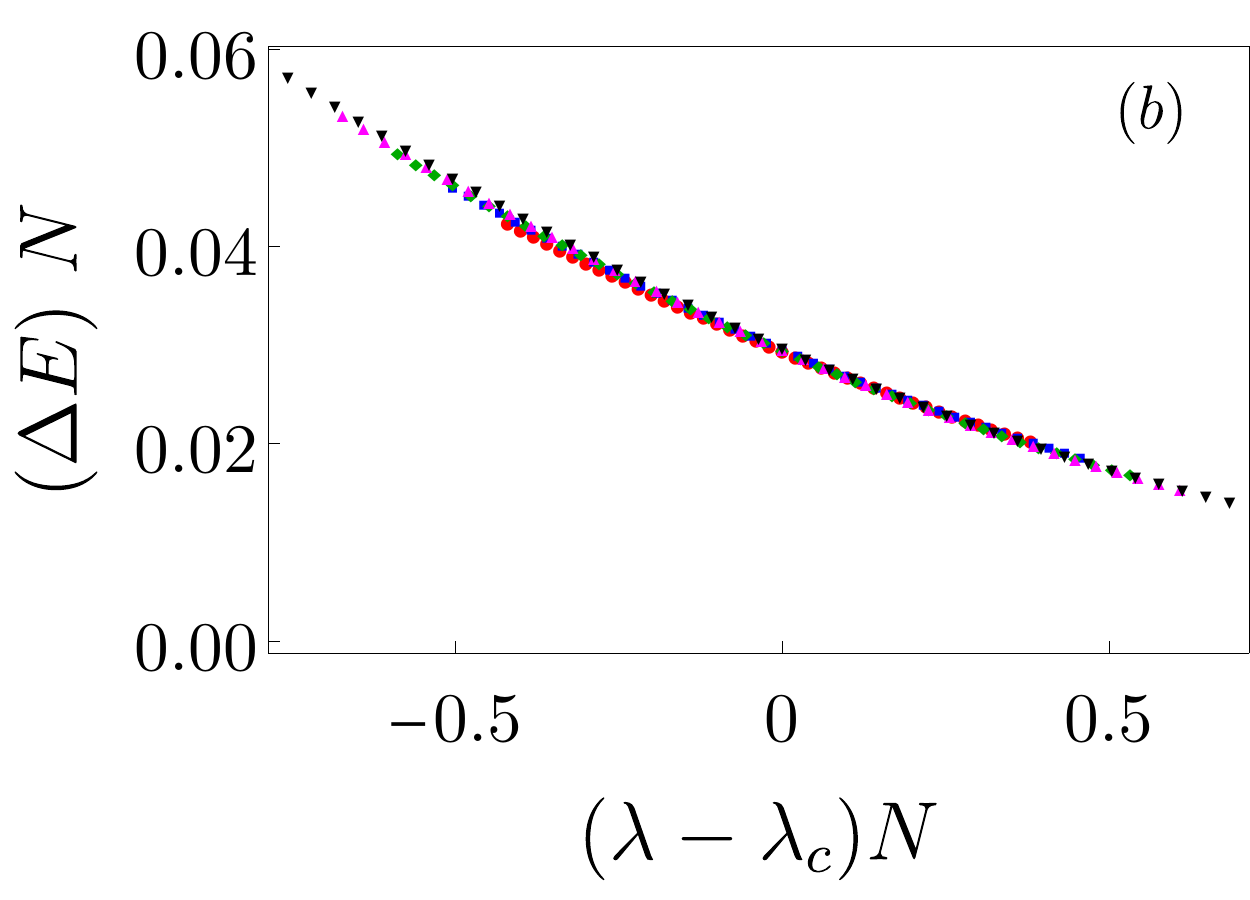}
\includegraphics[width=0.49\linewidth]{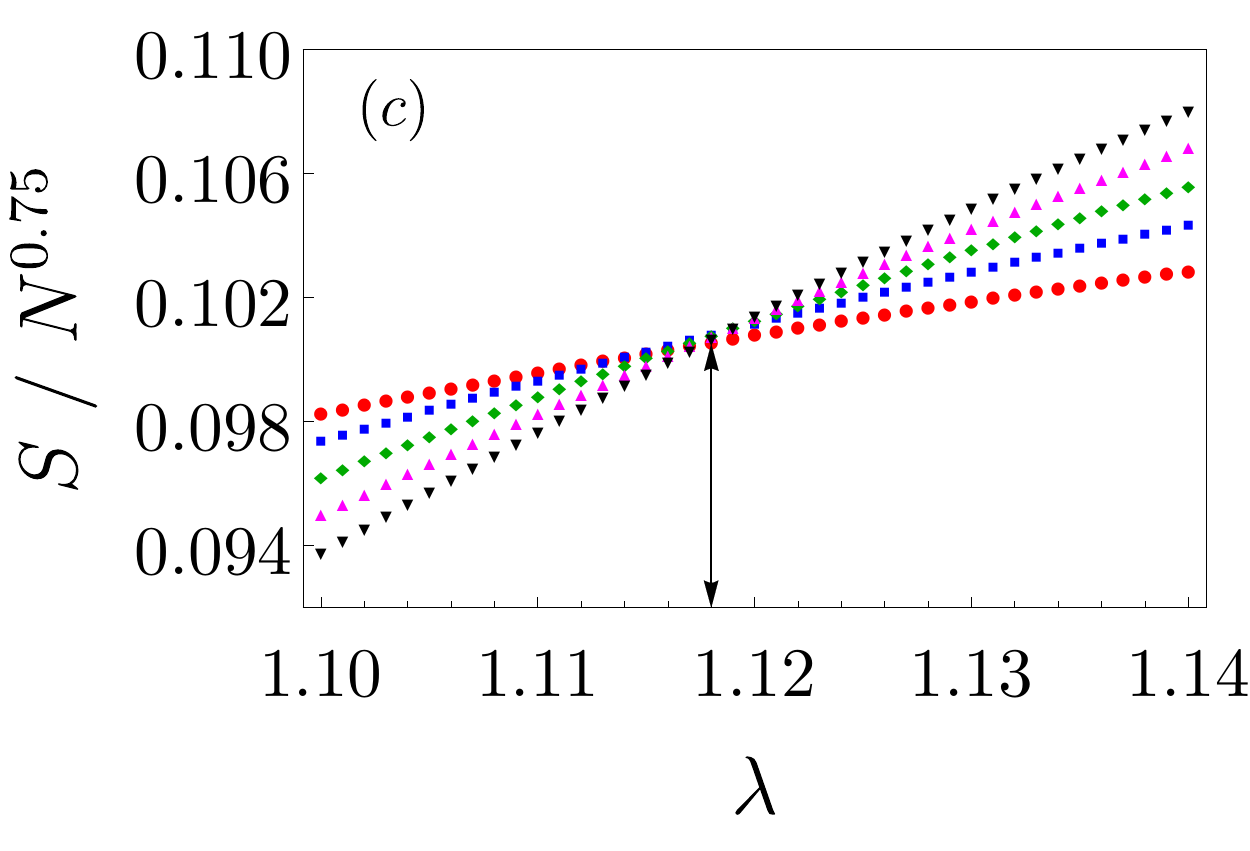}
\includegraphics[width=0.49\linewidth]{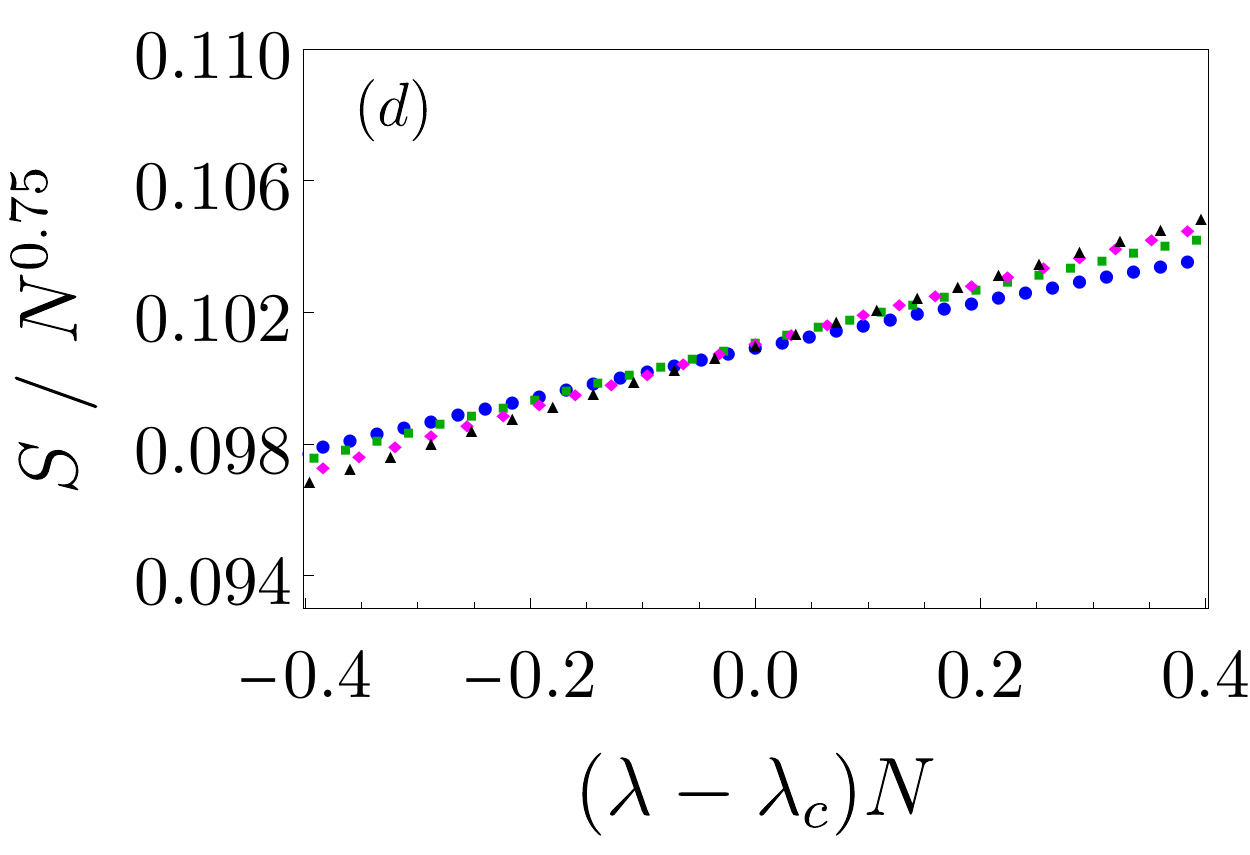}
\caption{(a): Exact numerics confirming the presence of the Ising
transition for $V_0=2.0$ and $w=0.1$ (a): Plot of $(\Delta E) N^z$
for a two-leg ladder as a function of $\lambda$ for several $N$
showing a crossing at $\lambda=\lambda_c \simeq 1.121$ for $z=1$.
(b) Plot of $(\Delta E) N^z$as a function of
$N^{1/\nu}(\lambda-\lambda_c)$ showing perfect scaling collapse for
$z=\nu=1$. (c): Plot of $S N^{2-z-\eta}$ as a function of $\lambda$
showing a crossing at $\lambda_c \simeq 1.118$ for $\eta=0.25$. (d)
Plot of $S N^{2-z-\eta}$ as a function of
$(\lambda-\lambda_c)N^{1/\nu}$ showing scaling collapse for $\nu=1$,
$\lambda_c \simeq 1.118$, and $\eta=0.25$. All energies are scaled
in units of $\Delta$. See text for details.} \label{fig6}
\end{figure}

\begin{figure}[ht]
\centering
\includegraphics[width=0.49\linewidth]{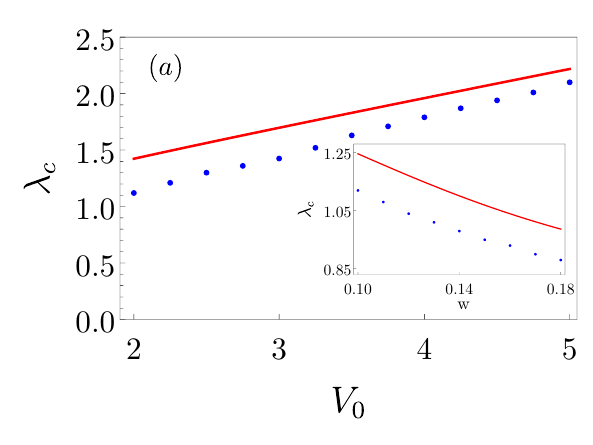}
\includegraphics[width=0.49\linewidth]{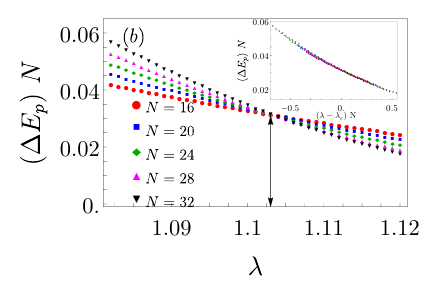}
\caption{(a): Plot of $\lambda_c$ as a function of $V_0$ for
$w_0=0.1$ showing the presence of the transition for a wide range of
$V_0 > V_c$. The inset shows a plot of $\lambda_c$ as a function of
$w$ for $V_0=2$. For both plots, the blue dots represent exact
results obtained using ED while the red lines shows the analytical
result obtained using Eq.\ \ref{vv6}. (b) A plot of $(\Delta E_p)
N^z$ as a function of $\lambda$ for different $N$ showing a crossing
at $\lambda_c \simeq 1.104$ for $z=1$. The inset shows a plot
$(\Delta E_p) N^z$ as a function of $(\lambda-\lambda_c)N^{1/\nu}$
showing data collapse for $z=\nu=1$. These plots indicate that the
Van-Vleck perturbation theory captures the essence of the Ising
transition. All energies are scaled in units of $\Delta$. See text
for details. } \label{fig7}
\end{figure}

Next, we carry out finite-size scaling analysis near the transition.
The results are indicated in Fig.\ \ref{fig6}. Our analysis reveals
the presence of a quantum critical point at $\lambda_c \simeq 1.12
\Delta$ for $V_0=2 \Delta$ and $w=0.1 \Delta$. The Ising
universality class of this transition is confirmed from finite-size
scaling as shown in Fig.\ \ref{fig6}; our results indicate $z=\nu=1$
and are consistent with $\eta=0.25$ even though the scaling collapse
of $S$ suggests significant finite-size scaling corrections for the
system sizes accessed using ED. The variation of $\lambda_c$ with
$V_0> V_c$ and $w$ is shown in Fig.\ \ref{fig7}(a). We note that the
transition exists for a wide range of $V_0$ and $w$; $\lambda_c$
varies linearly with both $V_0$ and $w$ within this range. Moreover,
we find that in contrast to its counterpart in $H_{\rm eff}$,
$\lambda_c$ depends on $w$ as shown in the inset of Fig.\
\ref{fig7}(a).

To obtain an analytic estimate of the shift in the position of the
transition, we use a Van-Vleck perturbation theory to construct an
effective low-energy Hamiltonian. To this end, we write $H^V_{\rm
eff} = H_0^V +H_1$ where $H_1 = w \sum_{j=1}^{2L} \sum_{\ell=1}^{2}
\tilde \sigma^x_{j,\ell}$ and $H_0^V$can be read off from Eq.\
\ref{zeroproj1}. We first identify the low-energy manifold of states
corresponding to $H^V_0$; these states are represented as
$|m\rangle$. Next, we construct a canonical transformation $S'$
leading to an effective Hamiltonian
\begin{eqnarray}
H'_V &=& e^{iS'} H_{\rm eff}^V e^{-iS'} = H_0^V + H_1 +[iS',
H_0+H_1] + \frac{1}{2} [iS',[iS',H_0^V]] + .., \label{vv1}
\end{eqnarray}
where the ellipsis indicate higher order terms in $w$ which we
neglect. Note that the presence of the interaction term in $H_0^V$
makes it difficult to obtain a closed form operator expression for
$S'$; however, its matrix elements can still be analytically
expressed. To determine $S'$, we again demand that it is chosen such
that all ${\rm O}(w)$ terms, which takes one out the low-energy
manifold of states $|m\rangle$, are eliminated; this leads to matrix
elements of $S'$ as
\begin{eqnarray}
\langle n|iS'|m\rangle &=& \frac{ \langle n
|H_1|m\rangle}{E_n^0-E_m^0}, \quad  \langle m|iS'|m'\rangle =0,
\label{vv2}
\end{eqnarray}
where $|n\rangle$ denotes eigenstates of $H_0^V$ which are not part
of the low-energy manifold and $E_n^0$ are the corresponding
eigenvalues. We also note that the second relation in Eq.\ \ref{vv2}
is automatically satisfied when $w$ is the large compared to energy
width of the low-energy manifold (see Fig.\ \ref{fig5}(d)); this is
the regime which we shall be interested in for the rest of this
section. One can then substitutes Eq.\ \ref{vv2} in Eq.\ \ref{vv1}
and obtain the matrix elements of $H'_V$ within the low-energy
manifold
\begin{eqnarray}
\langle m|H'_V|m'\rangle &=& E_m^0 \delta_{mm'} - \frac{w^2}{2}
\sum_{n} \langle m| \sum_{j=1}^{2L} \sum_{\ell=1}^{2} \tilde
\sigma_{j,\ell}^x |n\rangle \langle n| \sum_{j=1}^{2L}
\sum_{\ell=1}^{2}\tilde \sigma_{j,\ell}^x |m'\rangle \nonumber\\
&& \times \left(\frac{1}{E_n^0-E_m^0} +
\frac{1}{E_n^0-E_{m'}^0}\right), \label{vv3}
\end{eqnarray}
where the sum over $n$ indicates sum over states outside the
low-energy manifold. Here $E_m^0$ denotes the energy of the state
$|m\rangle$ due to $H_0^V$ while the terms $\sim w^2$ results from
second-order virtual processes due to $H_1$. We identify the
low-energy manifold of states numerically and diagonalize $\langle
m|H'_V|m'\rangle$ for several $N$. The resultant energy gap between
the ground and the first excited state is denoted by $\Delta E_{p}$;
a plot of $(\Delta E_p) N^z$ as a function of $\lambda$ is shown in
Fig.\ \ref{fig7}(b) for $z=1$. We find that it indicates the
presence of quantum phase transition belonging to Ising universality
at $\lambda_c \simeq 1.104 \Delta$; the inset of Fig.\ \ref{fig7}(b)
shows the plot of $(\Delta E_p) N^z$ as a function of
$(\lambda-\lambda_c)N^{1/\nu}$ showing scaling collapse for
$\lambda_c \simeq 1.104$ for $z=\nu=1$. These results agree
remarkably well with ED and shows that the essence of the Ising
transition is well captured by the perturbation theory.  We note
that when $V_{{\bf r} {\bf r'}}$ for $|{\bf r}-{\bf r'}|>1$ are
neglected, $E_m^0$ becomes a $m$ independent constant leading to an
exactly degenerate manifold with all states having $n=1/4$. The
presence of a finite  $V_{{\bf r} {\bf r'}}$ for $|{\bf r}-{\bf
r'}|>1$ lifts this degeneracy and also allows the ground state
manifold near the transition to have a small admixture of states
with $n>1/4$ in the presence of finite $w$ near the transition.

The picture of the transition that emerges from the
above analysis is as follows. We note that Fig.\ \ref{fig5}(d) shows
clear indication of near-degenerate low-energy manifold of states.
The presence of a finite $w$ leads to a ground state which is an
admixture of these low-energy nearly-degenerate manifold of states
arising out of a quantum order-by-disorder mechanism. This leads to
a unique ground state below a critical $\lambda =\lambda_c$; in
contrast, for $\lambda>\lambda_c$, the ground state breaks a
${\mathbb Z}_2$ symmetry which necessitates the presence of an Ising
critical point at $\lambda_c$.

To obtain an analytical estimate of the position of the critical
point, we now approximate the ordered ground state for $\lambda
>\lambda_c$ to be given by $|\psi_G\rangle$. This is not strictly
correct since numerically we find that the ordered state has a small
admixture of states with higher $n$. However, since this admixture
is small, the estimate we obtain using this approximation is
expected to be qualitatively correct. We then create an excited
state given by $|\psi_{\rm ex}\rangle$ (Eq.\ \ref{exwav}).  Since
both $|\psi_G\rangle$ and $|\psi_{\rm ex}\rangle$ are part of
low-energy manifold of states, we use Eq.\ \ref{vv3} to compute
$\langle H \rangle$ for them. A straightforward computation, keeping
terms  up to fourth nearest-neighbors in interaction ($|{\bf r}-{\bf
r'}| \le \sqrt{5}$), shows that
\begin{eqnarray}
E_{\rm ex}^V &=& \langle \psi_{\rm ex} |H'_v|\psi_{\rm ex}\rangle =
E_{G}^V + \frac{\delta E_1^V}{2}+\delta E_2^V, \quad  E_{G}^V= \langle \psi_G |H'_v|\psi_G\rangle, \nonumber\\
E_G^v &=& L \left( -(\Delta+\lambda) + \frac{V_0}{2^6}
-\frac{w^2}{(\Delta-\lambda) + V_0/4} \right) +E_0 \nonumber\\
\delta E_1^V &\simeq& 2 V_0 n_1  + \frac{w^2}{\frac{V_0}{4}+(\Delta
-\lambda)}, \quad \delta E_2^V \simeq -\frac{w^2}{\Delta+\lambda},
\quad n_1= \left(\frac{1}{(\sqrt{5})^6 }-\frac{1}{2^6}\right),
\label{vv4}
\end{eqnarray}
where $E_0$ is the energy of the all-spin down state and we have
neglected terms ${\rm O}(w^2 V_0$ \\ $/(2^6 (\Delta+\lambda)^2))$
and ${\rm O}(w^2 V_0/((\sqrt{5})^6 (\Delta+\lambda)^2))$; these
terms are always small in the range of $V_0$ and $w$ we are
interested in. The virtual processes responsible for the $O(w^2)$
term turns out to be same that in Fig.\ \ref{fig4}; their
amplitudes, however, change due to the presence of $V_0$. We note
that for $w=0$, $\delta E_{\rm ex} <0$ which means the ordered phase
never occurs in absence of the order-by-disorder mechanism realized
via ${\rm O}(w^2)$ terms in $H'_V$. Second, for $V_0=0$, $\delta
E_{1(2)}^V = \delta E_{1(2)}$ and we get back Eq.\ \ref{encal} for
the excitation energy. Equating $\delta E_{\rm ex}^V =0$, we finally
find
\begin{eqnarray}
\lambda_c &=& \frac{ V_0^2 n_1 + 6 w^2 - \sqrt{(V_0 n_1 (8\Delta
+V_0) -2 w^2)^2 +32 w^4}}{8 V_0 n_1}. \label{vv6}
\end{eqnarray}
We note that $V_0=2 \Delta$ and $w=0.1 \Delta$, Eq.\ \ref{vv6}
yields $\lambda_c \simeq 1.25 \Delta$ which is close to the exact
result obtained from ED: $\lambda_c^{\rm exact}=1.12 \Delta$.  A
plot of $\lambda_c$  as a function of $V$ and $w$ shown in Fig.\
\ref{fig7}(a). Comparting the analytical result (red solid line)
with the numerical ones (blue dots), we find that the former
overestimates $\lambda_c$ by a small amount but provides similar
dependence of $\lambda_c$ on both $V$ and $w$. The numerical
difference between the two arises from the perturbative nature of
the Van-Vleck result and also from the approximate nature of the
ordered ground state since we have neglected the small admixture
with states having $n>1/4$.  We note that the presence of a critical
lower value of $w$ is clearly required for $\lambda_c <\lambda_u$;
for $w<w_c$, which can be estimated by equating
$\lambda_c=\lambda_u$ in Eq.\ \ref{vv6}, the transition does not
occur.

Thus our analysis shows that the presence of a finite $V_0$ does not
obliterate the Ising transition. This allows for the possibility of
concrete realization of such a transition in experimentally relevant
Rydberg atom systems. Importantly, the perturbative analysis clearly
indicates that the transition is stabilized by a quantum
order-by-disorder mechanism.

\section{Discussion}
\label{diss}

In this work we have identified a quantum phase transition which is
stabilized by a quantum order-by-disorder mechanism in Rydberg
ladders with staggered detuning. Such a transition does not have any
analogue in Rydberg chains studied earlier. We have shown that this
transition persists for a wide range of vdW interaction strength
between the Rydberg atoms. Our numerical studies are supplemented by
perturbative calculations which provides analytical insight into the
nature of the transition via identification of the competing terms
in the low-energy effective Hamiltonian of the system responsible
for it; moreover, they provide remarkably accurate estimate of the
critical detuning.

Our prediction can be tested using standard experiments on Rydberg
atom systems \cite{rydexp1,rydexp2}. The simplest geometry to
consider would be the two-leg ladder with $w\ll |\lambda|, \Delta$
and a finite $V_0$ within range shown in Fig.\ \ref{fig7}(a). We
envisage a situation where the detunings at the odd(even) sites of
the Rydberg chains, given by $\lambda_{\rm odd}(\lambda_{\rm
even})$, are \cite{siteexp}
\begin{eqnarray}
\lambda_{\rm even} &=& \lambda+\Delta, \quad \lambda_{\rm odd}=
\lambda-\Delta. \label{exp1}
\end{eqnarray}
The Ising transition is then expected to occur, for example, for
$V_0=3 \Delta$, at $\lambda_c \simeq 1.5 \Delta$ which translates to
$\lambda_{\rm even}/\lambda_{\rm odd} \simeq 5$. These values can of
course be tuned by appropriate tuning of $V_0$ and $w$ within the
specified range discussed in Sec.\ \ref{pot}. In the ordered phase,
with $\lambda>\lambda_c$, all the Rydberg excitations will
preferentially happen on one of the two ladder provided one breaks
the $\mathbb{Z}_2$ symmetry by changing the detuning slightly on any
even site of the one of the chains. In contrast, the Rydberg
excitations in the same setting will be distributed on both chains
for $\lambda_{\rm even}/\lambda_{\rm odd} <5$. This change should be
easily detected by standard fluorescent imaging techniques used for
detection of Rydberg excitations in these systems
\cite{rydexp1,rydexp2,rydexp3,rydexp4}.

Finally, our analysis of the Rydberg ladder in the limit $V_{\bf r
\bf r'} \ll w, \lambda, \Delta$ points towards an interesting class
of spin models where $\mathbb{Z}_{\ell_0}$ symmetry broken ground
states and associated quantum phase transitions are stabilized by an
order-by-disorder mechanisms. We note that this is in contrast to
other spin models where a critical point arises due to competition
between a classical term and an order-by-disorder induced term in
the Hamiltonian \cite{spinref}. For a ladder with $\ell_0$ chains,
as shown in App.\ \ref{multileg}, a $\mathbb{Z}_{\ell_0}$ symmetry
broken ground state is realized for $\lambda
> \lambda_c= \Delta(\ell_0-1)/(\ell_0+1)$; in contrast, for $\lambda<\lambda_c$,
the ground state does not break any symmetry. This leads to the
realization of a critical point at $\lambda=\lambda_c$ which, for
example, belongs to three-state Potts university class for
$\ell_0=3$. However, unlike $\ell_0=2$, ladders with $\ell_0 >2$
chains also require periodic boundary conditions along the rung
direction to stabilize non-Ising transitions; this might be
difficult to achieve in current experimental setups. In conclusion
we have studied the phase diagram of Rydberg ladders with $\ell_0$
legs in the presence of staggered detuning. Our study indicates that
the low-energy behavior of such systems is described by class of
constrained spin models which support $\mathbb{Z}_{\ell_0}$ symmetry
broken ground states and associated emergent quantum criticality
stabilized by an order-by-disorder mechanism. For the two-leg
ladder, we show, by considering realistic vdW interactions, that the
presence of such an interaction do not obliterate the emergent Ising
transition; this leads to the possibility of its detection in
standard experimental setup.

\section*{Acknowledgement}

MS thanks Roopayan Ghosh for discussions. KS thanks DST, India for
support through SERB project JCB/2021/000030.

 \nolinenumbers

\appendix

\section{Variational wavefunction approach for $\lambda \simeq
\Delta$} \label{varwav1}

In this section, we shall address the ground state of the system
around $\lambda=\Delta$ and for $\lambda, \Delta \gg w$. To this
end, we first note, that for $\lambda \gg \Delta, w$, the ground
state of the system prefers maximal number of Rydberg excitations
and can therefore be written as
\begin{eqnarray}
|\psi_3 \rangle &=& \cos \phi_1 |\psi_{3a}\rangle + \sin
\phi_1 |\psi_{3b} \rangle \nonumber\\
|\psi_{3a} \rangle &=& \prod_{j=1}^{L/2}
|\uparrow_{2j-1,1},\downarrow_{2j-1,2};
\downarrow_{2j,1},\uparrow_{2j,2}\rangle, \quad  |\psi_{3b}\rangle =
|\downarrow_{2j-1,1},\uparrow_{2j-1,2};
\uparrow_{2j,1},\downarrow_{2j,2}\rangle, \label{tlstate1}
\end{eqnarray}
where once again we have chosen $\phi_1$ to be independent of $j$.
In contrast, for $\lambda \le \Delta$, the ground state prefers to
have down-spins on the odd sublattices. Thus one may choose the
variational ground state wavefunction to be $|\psi_4(\phi_2)\rangle
\equiv |\psi_4\rangle = |\psi_2(\phi=\phi_2)\rangle$ where
$|\psi_2\rangle$ is given by Eq.\ \ref{state2}.  Using these
wavefunctions, one can therefore construct a variational
wavefunction given by
\begin{eqnarray}
|\Phi_v\rangle &=& \cos \beta |\psi_3\rangle + \sin \beta |\psi_4
\rangle, \label{varpos2leg}
\end{eqnarray}
where $ 0\le \beta \le \pi/2$. The corresponding variational energy
can be computed in a straightforward manner and yields
\begin{eqnarray}
E_{1v} &=& \langle \Phi_v| H_2 |\Phi_v\rangle  =  E_0 +
(\lambda-\Delta) \cos^2 \beta - w \sin(2\beta)
\cos(\phi_1-\phi_2+\pi). \label{varen2leg}
\end{eqnarray}
The minimization of $E_{1v}$ leads to
\begin{eqnarray}
\phi_1 &=& \phi_2+\pi, \quad \beta_0 = \frac{1}{2} \arctan
\left[\frac{2 w}{\Delta-\lambda}\right]. \label{parsol2leg}
\end{eqnarray}

Next, we consider the change in expectation values of the operators
$\hat n$ and $s_x$ across the transition. To this end, we note that
using Eqs.\ \ref{varpos2leg} and \ref{parsol2leg}, we find
\begin{eqnarray}
\langle \hat n \rangle_v &=& \frac{1}{2}-\frac{1}{4} \cos^2 \beta_0
=\frac{1}{2}-\frac{1}{8}
\left(1-\frac{\lambda-\Delta}{\sqrt{(\lambda-\Delta)^2+4
w^2}}\right), \nonumber\\
\langle s'_x \rangle_v &=& \frac{1}{N} \langle \sum_{j=1}^L
\sum_{\ell=1}^{2} \langle \sigma^x_{2j-1, \ell} \rangle= \frac{\sin
2\beta_0}{4}= \frac{w}{2\sqrt{(\lambda-\Delta)^2+4 w^2}}.
\label{varex}
\end{eqnarray}
These expectations values, can also be computed using the exact
ground state obtained from ED. The two results show excellent match
similar to their counterparts near $\lambda \sim -\Delta$ discussed
in the main text. The plot for $n$ near $\lambda \sim \Delta$, shown
in Fig.\ \ref{fig1}(a), corroborates the above statement. We note
here that the two states $|\psi_3\rangle$ and $|\psi_4\rangle$
belong to two different $\mathbb{Z}_2$ orders corresponding to
$n=1/4$ and $n=1/2$ respectively. Thus it is natural to expect a
first order transition between them around $\lambda \simeq \Delta$.
However, our numerics from ED seems to indicate a crossover which
can be inferred from $N$ independence of the data. This is most
likely a consequence of the presence of other sector states with
fixed $n$. These states do not feature in this simple minded
variational wavefunction computation and their presence may change
this transition to a crossover.

\section{$\ell_0$-leg ladder}
\label{multileg}

\begin{figure}[ht]
\centering
\includegraphics[width= \linewidth]{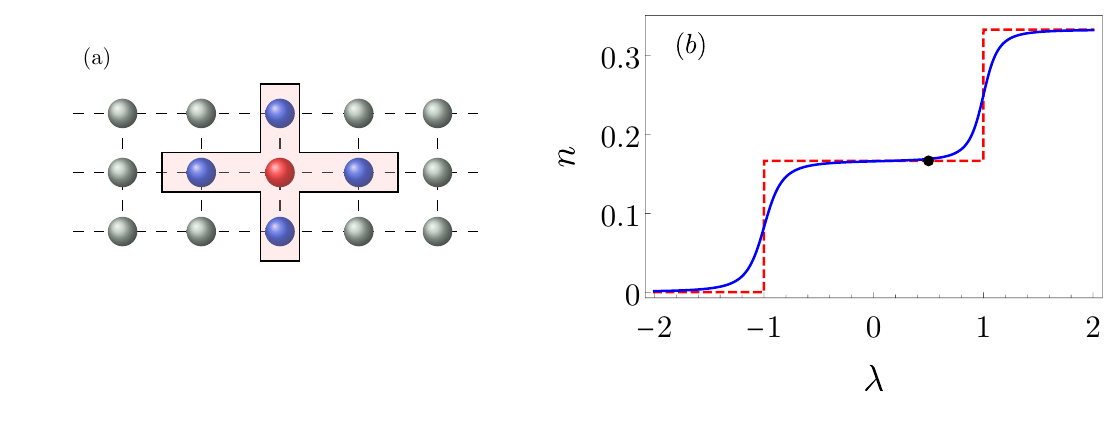}
\caption{(a) Schematic representation of the Rydberg ladder for
$\ell_0=3$ showing the Rydberg blockade radius (rectangular region)
around a Rydberg excitation (red circle). The blue circles denote
sites with blocked Rydberg excitation. The black circle at $\lambda
\sim 0.5$ shows the position of the quantum phase transition.(b) The
plot of Rydberg excitation density $n$ for a $\ell_0=3$ leg ladder
as a function of $\lambda$. The red dotted line corresponds to $w=0$
showing first order transition between the plateaus. The blue solid
lines correspond to $w=0.05$. The black circle shows the position of
the quantum phase transition. All energies are scaled in units of
$\Delta$. See text for details.} \label{figapp1}
\end{figure}

In this section, we consider the model $H_{\rm eff}$ for
$\ell_0$-legged ladder. For $\ell_0 >2$, such ladders are difficult
to realize in an experimentally relevant Rydberg atom arrays since
periodic boundary conditions is required along the rung direction.
However we study such model Hamiltonians here since they support
phase transitions with non-Ising universality that are stabilized by
order-by-disorder mechanism. In what follows, we shall provide a
general expression for the low-energy effective Hamiltonian
$H'_{\ell_0}$ for a ladder with $\ell_0$ legs. We shall also provide
numerical evidence of the transition for $\ell_0=3$ legged ladder.

To this end, we consider an arrangement of $\ell_0$ Rydberg chains
each having $2L$ sites so that $N=2 L\ell_0$ as shown in Fig.\
\ref{figapp1}(a) for $\ell_0=3$. As shown in this figure, we
consider a situation where the presence of a Rydberg excited atom on
any site precludes the presence of another such excitation in all of
the $\ell_0-1$ sites of the same rung of the ladder. The model
Hamiltonian that we shall consider for these chains is a
generalization for $H_{\rm eff}$ (Eq.\ \ref{effham1}) and is given
by
\begin{eqnarray}
H &=& \sum_{j=1}^{2L} \sum_{\ell=1}^{\ell_0} \left(w \tilde
\sigma_{j,\ell}^x - \frac{1}{2}[\Delta (-1)^j +
\lambda]\sigma_{j,\ell}^z \right), \quad  \tilde \sigma^x_{j,\ell} =
P_{j-1, \ell} \left(\prod_{\ell' \ne \ell} P_{j, \ell'}\right)
\sigma^x_{j,\ell} P_{j+1,\ell}. \label{twoham}
\end{eqnarray}

\begin{figure}[ht]
\centering
\includegraphics[width=0.49\linewidth]{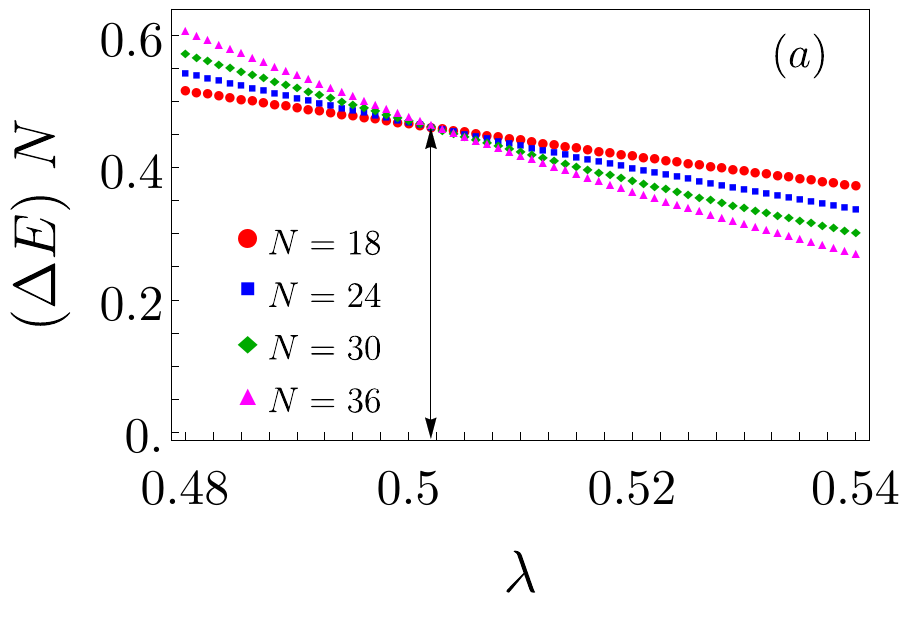}
\includegraphics[width=0.49\linewidth]{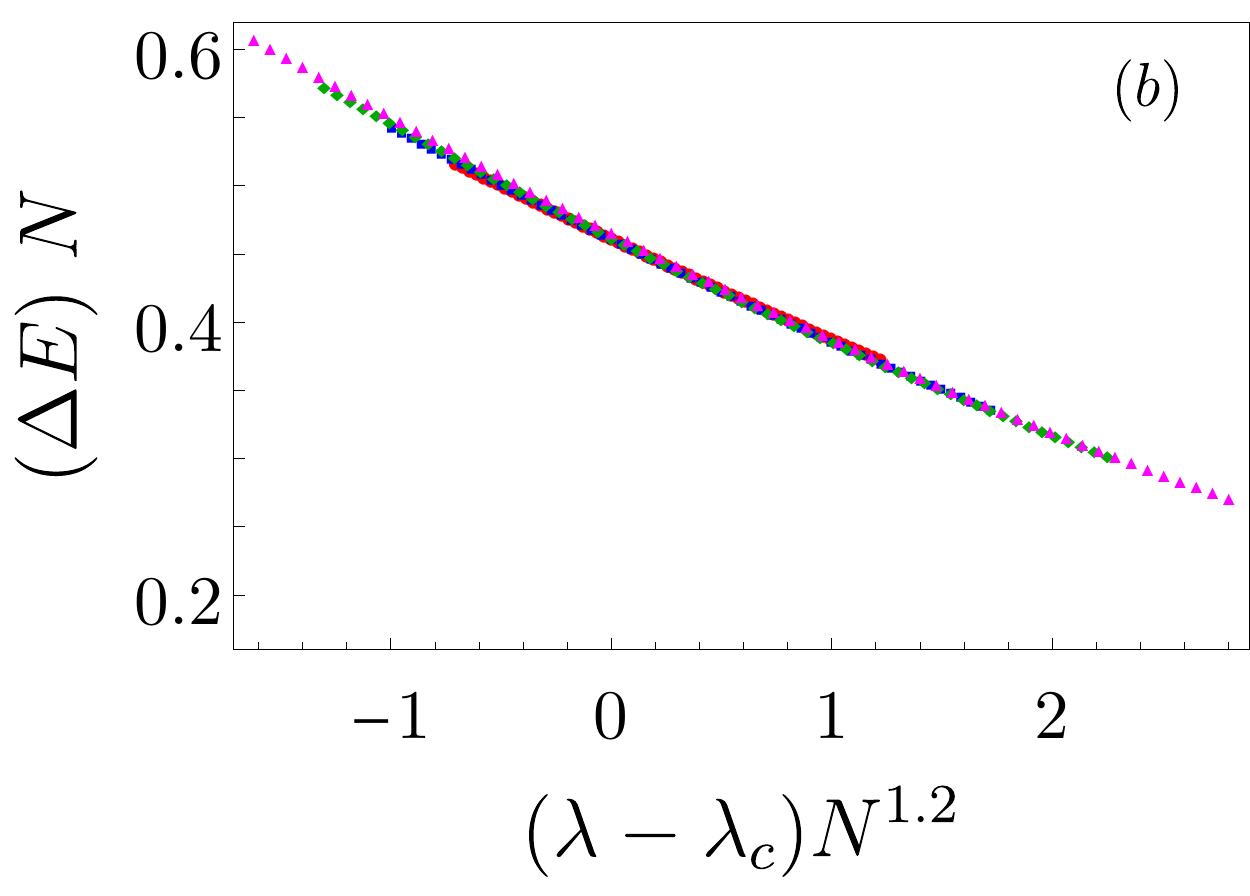}
\includegraphics[width=0.49\linewidth]{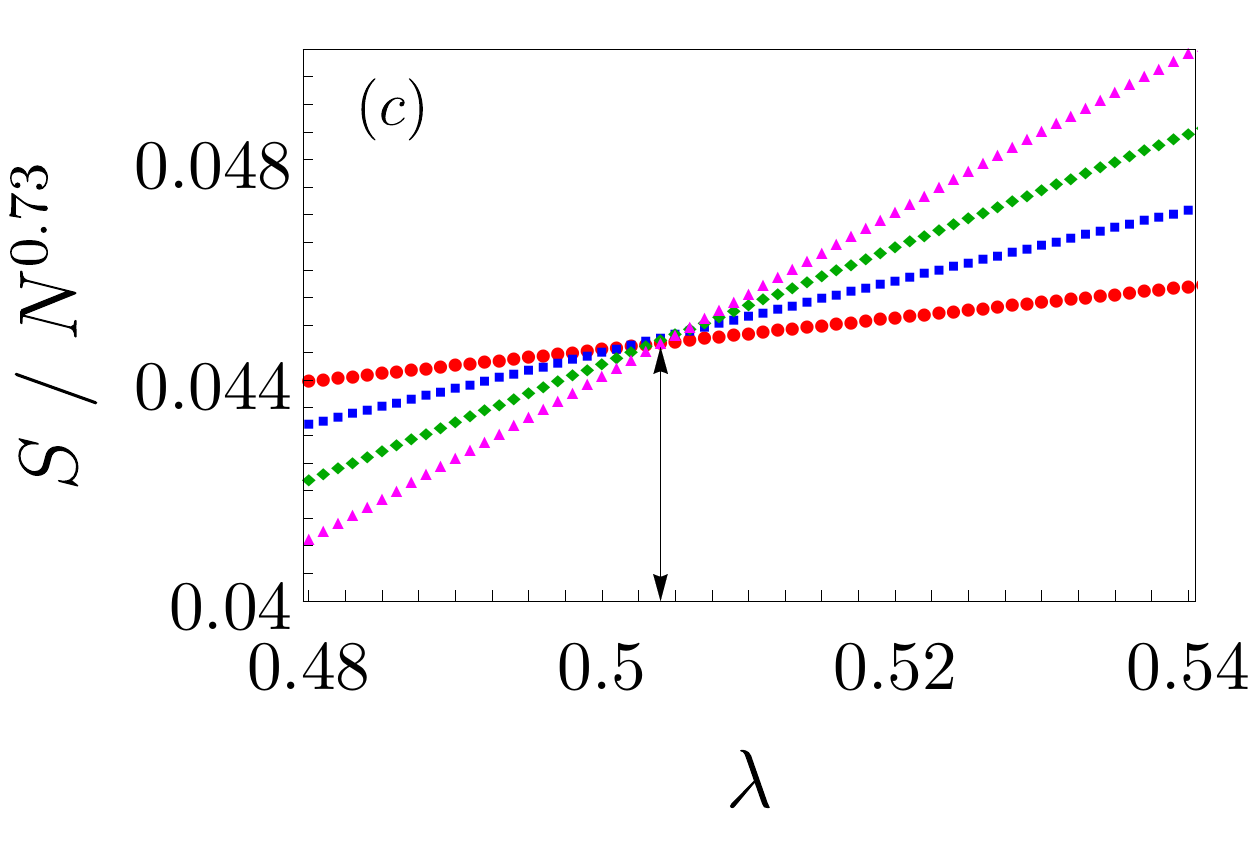}
\includegraphics[width=0.49\linewidth]{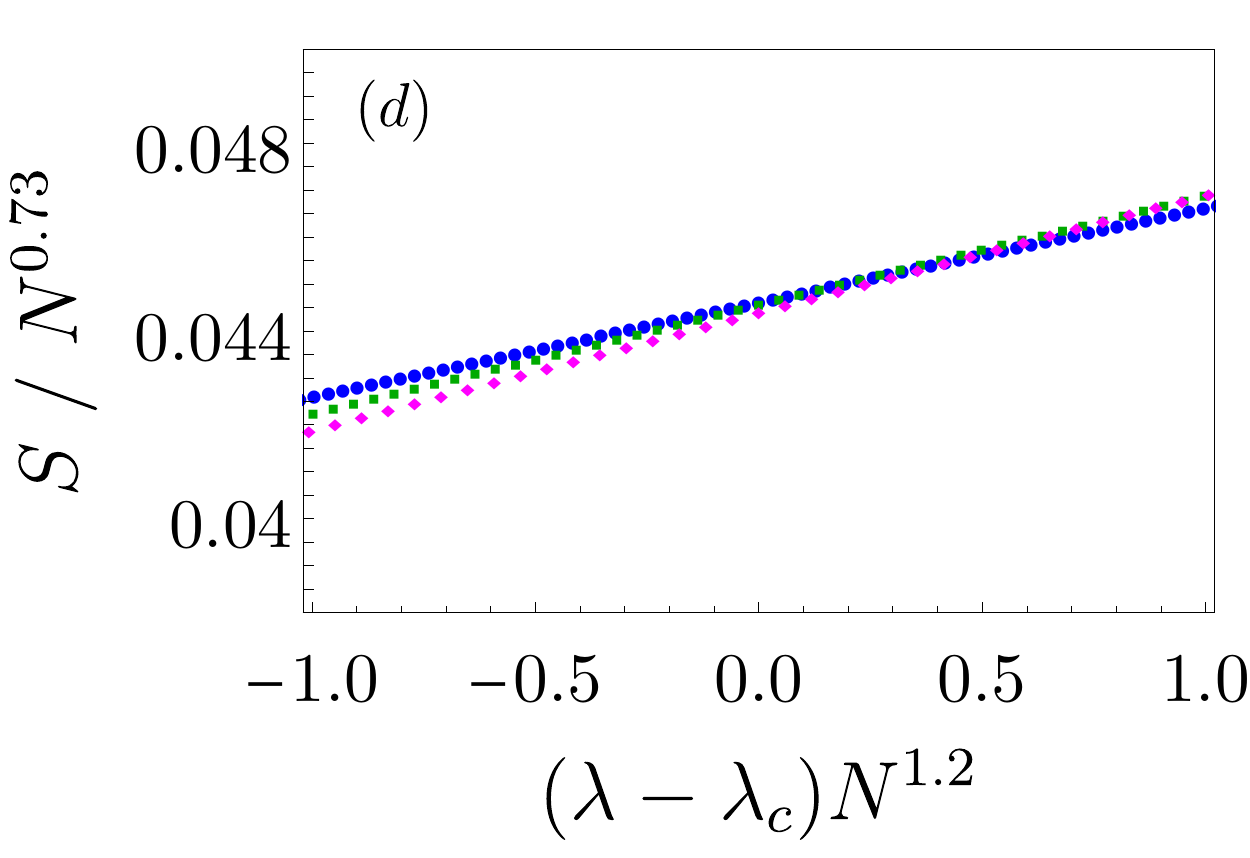}
\caption{ Exact numerics confirming the presence of the three-state
Potts transition for $\ell_0=3$. (a): Plot of $(\Delta E) N^z$ for a
three-leg ladder as a function of $\lambda$ for several $N$ showing
a crossing at $\lambda=\lambda_c \simeq 0.501$ for $z=1$. (b) Plot
of $(\Delta E) N^z$ as a function of $N^{1/\nu}(\lambda-\lambda_c)$
showing perfect scaling collapse for $z=1$ and $\nu \simeq 1/1.2=
0.83$.(c): Plot of $S N^{2-z-\eta}$ as a function of $\lambda$
showing a crossing at $\lambda_c \simeq 0.502$ for $\eta \simeq
0.27$. (d) Plot of $S N^{2-z-\eta}$ as a function of
$\lambda-\lambda_c)N^{1/\nu}$ showing a scaling collapse for
$\nu=1/1.2$ and $\eta=0.27$. All energies are scaled in units of
$\Delta$. See text for details.} \label{figapp2}
\end{figure}

The phase diagram for such chains for $w=0$ can be obtained in a
straightforward manner. For $\lambda<0$, and $|\lambda| \gg \Delta$,
$n=0$. Around $|\lambda| \simeq \Delta$, it becomes energetically
favorable to create one Rydberg excitation on one of the $\ell_0$
sites of an even rung. This leads to a classical ground state with
$n_0=1/(2\ell_0)$; the ground state is macroscopically degenerate
since there are $\ell_0^L$ Fock states with $n=1/(2\ell_0)$. For
$\lambda >\Delta$, the ground state corresponds to maximum possible
Rydberg excitations leading to $n=1/\ell_0$; it exhibits a $\ell_0$
fold degeneracy. For $w=0$, there are first-order transitions
between these ground states at $\lambda/ \Delta \simeq \pm 1$; for
finite $w$, these are related by crossovers. The representative
phase diagram, for $\ell_0=3$, is shown in Fig.\ \ref{figapp1}(b).
Our numerical analysis also detects, for $\ell_0=3$, a quantum
transition within the $n=1/6$ plateau at $\lambda_c/\Delta \simeq
0.52$. A finite-sized scaling analysis near of the energy gap
$\Delta E$, as shown in Fig.\ \ref{figapp2}(a) and \ref{figapp2}(b)
indicates $z=1$ and $\nu =5/6 \simeq 0.83$ for the transition.
Furthermore we define an order parameter $\hat O_{\ell_0} =
\sum_{j=1}^{L} \sum_{\ell=1}^{\ell_0} \exp[2\pi i \ell/\ell_0]
\sigma_{2j, \ell}^z$ and computes its correlation function for the
$\ell_0=3$ chain: $S = \langle \hat O_3^{\ast} \hat O_3 \rangle/N$.
We find that, using a finite sized scaling analysis, $S \sim N^{2-z
-\eta}$ with $\eta=4/15 \simeq 0.267$ and $z=1$ as shown in Figs.\
\ref{figapp2}(c) and \ref{figapp2}(d). Thus we conclude that the
transition belongs to the 2D three-state Potts universality class.

An analytic insight into this transition is provided by a
calculation similar to that outlined in the main text leading to
Eq.\ \ref{effproj}. For an $\ell_0$-leg the effective Hamiltonian
receives contribution from processes which are similar to those
shown in Fig.\ \ref{fig4} (generalized for $\ell_0$ legs) and is
given by \cite{conf1,ksen1}
\begin{eqnarray}
&& H_{\rm eff}^{\ell_0} = \frac{-w^2}{\Delta-\lambda} \sum_{j=1}^{L}
\Big[ \sum_{\ell=1}^{\ell_0} P_{2j, \ell} P_{2j+2,\ell} +
\frac{\alpha_0}{2} \sum_{\ell,\ell'=1,\ell_0 ; \ell \ne \ell'}
P_{2j, \ell'} (\sigma^x_{2j, \ell} \sigma^x_{2j, \ell'}
+\sigma^y_{2j, \ell} \sigma^y_{2j, \ell'} )\Big], \nonumber\\
\label{effproj2}
\end{eqnarray}
where $\alpha_0= (\Delta-\lambda)/(\Delta+\lambda)$. The emergence
of the phase transition at a critical value of $\lambda/\Delta$ can
once again be understood by noting the competition between the two
terms in Eq.\ \ref{effproj2}. The first term prefers a $\mathbb{Z}_{\ell_0}$
symmetry broken ground state with all Rydberg excitations on any one
of the $\ell_0$ chains; in contrast, the second term prefers a
linear superposition of all states with one Rydberg excitation on
any site of even rungs of the ladder. This necessitates an
intermediate quantum critical point belonging to $\mathbb{Z}_{\ell_0}$
universality class.

To estimate the position of this critical point, we once again
consider an excited state over the $\mathbb{Z}_{\ell_0}$ symmetry broken
ground state $|\psi'_G\rangle$ given by
\begin{eqnarray}
|\psi'_G\rangle &=& |\downarrow_{1,1} \downarrow_{1,2}..
\downarrow_{1,\ell_0}; \uparrow_{2,1}
\downarrow_{2,2}..\downarrow_{2,\ell_0}; ....\uparrow_{2j,1}
\downarrow_{2j,2}..\downarrow_{2j,\ell_0}...;\uparrow_{2L,1}
\downarrow_{2L,2}.. \downarrow_{2 \ell_0}\rangle.
\end{eqnarray}
Such an excited state can be created by constructing a linear
superposition of states with a Rydberg excitation on any one of the
sites of one of the even rungs of the ladder for which $j'=2j$; for
all $j'\ne 2j$, the Rydberg excitation reside at $\ell_0=1$. Such an
excited state can be represented as
\begin{eqnarray}
|\psi'_{\rm ex}\rangle &=&
\frac{1}{\sqrt{\ell_0}}\left(|\downarrow_{1,1} \downarrow_{1,2}..
\downarrow_{1,\ell_0}; \uparrow_{2,1}
\downarrow_{2,2}..\downarrow_{2,\ell_0}; ....\uparrow_{2j,1}
\downarrow_{2j,2}..\downarrow_{2j,\ell_0}...;\uparrow_{2L,1}
\downarrow_{2L,2}.. \downarrow_{2 \ell_0}\rangle \right. \nonumber\\
&& \left. + |\downarrow_{1,1} \downarrow_{1,2}..
\downarrow_{1,\ell_0}; \uparrow_{2,1}
\downarrow_{2,2}..\downarrow_{2,\ell_0}; ....\downarrow_{2j,1}
\uparrow_{2j,2}..\downarrow_{2j,\ell_0}...;\uparrow_{2L,1}
\downarrow_{2L,2}.. \downarrow_{2 \ell_0}\rangle \right. \nonumber\\
&& \left. + .... |\downarrow_{1,1} \downarrow_{1,2}..
\downarrow_{1,\ell_0}; \uparrow_{2,1}
\downarrow_{2,2}..\downarrow_{2,\ell_0}; ....\downarrow_{2j,1}
\downarrow_{2j,2}..\uparrow_{2j,\ell_0}...;\uparrow_{2L,1}
\downarrow_{2L,2}.. \downarrow_{2 \ell_0}\rangle \right),
\nonumber\\
\label{exwavl}
\end{eqnarray}

The energy cost of creating such an excited state can be easily
computed following the method outlined in the main text and leads to
$\delta E_{\rm ex}^{\ell_0}= \delta E_1^{\ell_0} +\delta
E_2^{\ell_0}$ where
\begin{eqnarray}
\delta E_1^{\ell_0} &=& \frac{w^2(1-1/\ell_0)}{\Delta -\lambda},
\quad \delta E_2^{\ell_0} = -\frac{w^2 (\ell_0-1)}{\Delta+\lambda}.
\label{encall}
\end{eqnarray}
The energy gap vanishes when $\delta E^{\ell_0}_{\rm ex}=0$ which
leads to an estimate of the critical point given by
\begin{eqnarray}
\lambda_c = \Delta (\ell_0-1)/(\ell_0+1). \label{critesl}
\end{eqnarray}
For $\ell_0=2$, this reproduces Eq.\ \ref{crites} of the main text;
for $\ell_0=3$, it provide an estimate of the three-state Potts
critical point to be $\lambda_c=\Delta/2$ which matches quite well
with exact numerical results shown Fig.\ \ref{figapp2}.

Before ending this section, we note that we expect that for
$\ell_0>4$, the transition shall be first order; however we have not
been able to carry out ED studies on sufficiently large system size
to ascertain this due to increased dimension of the constrained
Hilbert space.

\bibliography{rydladsci_v7}

\end{document}